\documentclass[twocolumn]{aastex631}
\usepackage{comment}
\usepackage{enumitem}

\graphicspath{{./}{Figures/}}

\submitjournal{ApJ}

\shorttitle{Combined Study on LIV with Multiple-type Gamma-ray Sources: Part I}
\shortauthors{Bolmont et al.}

\usepackage{amsmath}

\usepackage{multirow}

\begin{document}

\title{First Combined Study on Lorentz Invariance Violation from Observations of \\
Energy-dependent Time Delays from Multiple-type Gamma-ray Sources\\
Part I - Motivation, Method Description and Validation through Simulations of \\ 
H.E.S.S., MAGIC and VERITAS Datasets}

\correspondingauthor{J. Bolmont}
\email{bolmont@lpnhe.in2p3.fr}

\author[0000-0003-4739-8389]{Julien Bolmont}
\altaffiliation{Author contributions are detailed in Appendix~\ref{sec:annexC}.}
\affiliation{Sorbonne Universit\'e, CNRS/IN2P3,\\ Laboratoire de Physique Nucl\'eaire et de Hautes Energies, LPNHE,\\ 4 Place Jussieu, F-75005 Paris, France}

\author[0000-0002-1103-130X]{Sami Caroff}
\altaffiliation{Present address: Laboratoire d’Annecy de Physique des Particules, Universit\'e Grenoble Alpes, Universit\'e Savoie Mont Blanc, CNRS, LAPP, F-74000 Annecy, France}
\affiliation{Sorbonne Universit\'e, CNRS/IN2P3,\\ Laboratoire de Physique Nucl\'eaire et de Hautes Energies, LPNHE,\\ 4 Place Jussieu, F-75005 Paris, France}

\author[0000-0001-8442-7877]{Markus Gaug}
\affiliation{Universitat Aut\`onoma de Barcelona and CERES-IEEC, E-08193 Bellaterra, Spain}

\author[0000-0001-7429-3828]{Alasdair Gent}
\affiliation{School of Physics \& Center for Relativistic Astrophysics,\\ Georgia Institute of Technology,\\ 837 State Street NW, Atlanta, GA 30332-0430}

\author{Agnieszka Jacholkowska}
\altaffiliation{Deceased.}
\affiliation{Sorbonne Universit\'e, CNRS/IN2P3,\\ Laboratoire de Physique Nucl\'eaire et de Hautes Energies, LPNHE,\\ 4 Place Jussieu, F-75005 Paris, France}

\author[0000-0002-5289-1509]{Daniel Kerszberg}
\affiliation{Institut de F\'isica d'Altes Energies (IFAE),\\ The Barcelona Institute of Science and Technology (BIST), E-08193 Bellaterra (Barcelona), Spain}

\author{Christelle Levy}
\affiliation{Sorbonne Universit\'e, CNRS/IN2P3,\\ Laboratoire de Physique Nucl\'eaire et de Hautes Energies, LPNHE,\\ 4 Place Jussieu, F-75005 Paris, France}
\affiliation{LUTH, Observatoire de Paris, PSL Research University,\\ CNRS, Universit\'e de Paris,\\ 5 Place Jules Janssen,
F-92190 Meudon, France}

\author{Tony Lin}
\affiliation{Physics Department,\\ McGill University,\\Montreal, QC H3A 2T8, Canada}

\author[0000-0002-9763-9155]{Manel Martinez}
\affiliation{Institut de F\'isica d'Altes Energies (IFAE),\\ The Barcelona Institute of Science and Technology (BIST), E-08193 Bellaterra (Barcelona), Spain}

\author{Leyre Nogu\'es}
\affiliation{Institut de F\'isica d'Altes Energies (IFAE),\\ The Barcelona Institute of Science and Technology (BIST), E-08193 Bellaterra (Barcelona), Spain}

\author[0000-0002-5955-6383]{A. Nepomuk Otte}
\affiliation{School of Physics \& Center for Relativistic Astrophysics,\\ Georgia Institute of Technology,\\ 837 State Street NW, Atlanta, GA 30332-0430}

\author{C\'edric Perennes}
\affiliation{Universit\`a di Padova and INFN,\\ I-35131 Padova, Italy}

\author[0000-0002-2160-2452]{Michele Ronco}
\affiliation{Sorbonne Universit\'e, CNRS/IN2P3,\\ Laboratoire de Physique Nucl\'eaire et de Hautes Energies, LPNHE,\\ 4 Place Jussieu, F-75005 Paris, France}

\author[0000-0002-4209-3407]{Tomislav Terzi\'c}
\affiliation{University of Rijeka, Department of Physics, 51000 Rijeka, Croatia}

\begin{abstract}


Gamma-ray astronomy has become one of the main experimental ways to test the modified dispersion relations (MDRs) of photons in vacuum, obtained in some attempts to formulate a theory of Quantum Gravity. The MDRs in use imply time delays which depend on the energy, and which increase with distance following some function of redshift. The use of transient, or variable, distant and highly energetic sources, already allows us to set stringent limits on the energy scale related to this phenomenon, usually thought to be of the order of the Planck energy, but robust conclusions on the existence of MDR-related propagation effects still require the analysis of a large population of sources. 

In order to gather the biggest sample of sources possible for MDR searches at teraelectronvolt energies, the H.E.S.S., MAGIC and VERITAS collaborations enacted a joint task force to combine all their relevant data to constrain the Quantum Gravity energy scale. In the present article, the likelihood method used, to combine the data and provide a common limit, is described in detail and tested through simulations of recorded data sets for a gamma-ray burst, three flaring active galactic nuclei and two pulsars. Statistical and systematic errors are assessed and included in the likelihood as nuisance parameters. In addition, a comparison of two different formalisms for distance dependence of the time lags is performed for the first time. In a second article, to appear later, the method will be applied on all relevant data from the three experiments.

\end{abstract}


\section{Introduction}

Modern physics is based on two fundamental pillars: quantum mechanics (QM) and Einsteinian general relativity (GR). When taken separately, these theories can claim success in satisfactorily describing many physical phenomena, but all attempts to make them compatible with each other have failed so far. The goal of quantum gravity (QG) research is to find a common approach to coherently merge quantum theory and GR. The QG problem has remained unsolved for more than eighty years now and keeps challenging physicists who, in the struggle to find a solution, have proposed a myriad of models  \citep[see e.g.][]{Polyakov1981,Bombelli1987,Oriti2006,Reuter2006,Rovelli2007,Loll2012}. However, none of these models can claim full success. One of the main obstructions to progress in this field is the lack of experimental guidance. However, in the last two decades, the situation has changed, and recent years have held important advances in the field of QG phenomenology \citep{Mattingly2005,Amelino2013,Liberati2013}. 

It is notoriously difficult to extract observable predictions from fully-fledged QG approaches. Different models usually start from different conceptual premises and use different mathematical formalisms in such a way it is difficult to determine whether they make compatible predictions. In some cases, the formal complexity forbids producing observable outcomes at all. Then, to guide experimental efforts, bottom-up approaches have been proposed \citep{Amelino2002,Kowalski2002,Smolin2004, Livine2011,Barrau2015,Ronco2018,Calcagni2019}. They rely on somewhat simpler models, suitable for describing only a subset of expected QG features, but have the advantage of producing opportunities for experimental testing. 

In this regard, at the end of the 90s, independent semi-classical analyses inspired by QG models brought to the attention of the QG community the fact that it is a highly non-trivial task to retain Lorentz symmetries when quantizing the space-time geometry of GR. These models include, most notably, String Theory \citep[see e.g.][and references therein]{Mavromatos2010}, Loop Quantum Gravity  \citep{Gamibini1999}, Non-commutative Geometry \citep{Carroll2001}, and Standard Model Extension \citep[][and references therein]{Kostelecky2008}. From then on, departures from Lorentz invariance have become one of the rare observable features we would expect in a QG theory and, as we shall see briefly, different bottom-up models to implement them have been proposed. According to this view, Lorentz invariance could be an emergent symmetry that arises in the low-energy limit but is modified at higher energies approaching the Planck scale, i.e. the energy scale at which both GR and QM effects should play an important role.

A much-studied way to encode departures from Lorentz invariance, either violations (noted LIV for Lorentz invariance violation) or deformations,
consists in modifying the energy-momentum dispersion relation of free relativistic particles as follows \citep{Amelino1998}: 

\begin{equation}
\label{eq:disprel1}
E^2 \simeq p^2 c^2\times\left[1 \pm \sum_{n=1}^\infty \left(\frac{E}{E_{QG}}\right)^n\right],
\end{equation}
where $c$ is (the low energy limit of) the speed of light, and $E_{QG}$ the energy scale of QG effects which is usually expected to be around the Planck scale ($E_P = \sqrt{\hbar c^5 / G}\, \simeq 10^{19}$ GeV).
The sign $\pm$ in Equation~(\ref{eq:disprel1}) takes into account the possibility to have subluminal or superluminal effects.

Published one year after the first redshift of a gamma-ray burst (GRB) was measured, the article by \citet{Amelino1998} also proposed for the first time the use of transient, distant and high-energy gamma-ray sources as a way to probe the quantum nature of space-time by searching for energy-dependent delays. In the following, we will focus on this particular way to probe a dispersion relation such as the one of Equation~(\ref{eq:disprel1}) in the so-called 
`time of flight' studies. Since then, other possibilities have emerged to search for QG effects in gamma-ray astronomy. For example, astrophysical sources have been used to search for vacuum birefringence \citep{Gotz2014}, and space-time `fuzziness' \citep{Vasileiou2015}. Possible modifications of the cross section of $\gamma\gamma$ interaction between high-energy photons and the extra-galactic background light were also investigated \citep{Biteau2015,Abdalla2019}. {Some of the limits published in these papers exceed the Planck scale, sometimes even by several orders of magnitude, but there is also a possibility that LIV could occur only through energy-dependent delays.} In principle, all these effects could {also} coexist, even if they have only been tested separately so far. {A comprehensive review of different possible effects of LIV on gamma rays, as well as on other messengers (cosmic rays, gravitational waves, neutrinos) is given by \citet{Addazi2021}.}

Heuristically, Equation~(\ref{eq:disprel1}) can be justified as follows: at Planckian distances ($\sim10^{-33}$ cm), QG effects are believed to cause fluctuations of space-time geometry which, then, would behave as a dynamical medium characterised by a non-trivial refractive index. Consequently, photons with different energies would have different interactions with the `foamy' structure of space-time (sometimes called `quantum space-time') and, thus, they would propagate in vacuum at different velocities thereby producing an effect of in-vacuo dispersion. This explains the dependence of Equation~(\ref{eq:disprel1}) on some power~$n$ of the energy $E$ of the probe. For simplicity,~$n$ is generally assumed to be an integer, and we will keep that assumption in this paper. However, in so-called fractional or (multi-)fractional models the modifications of the dispersion relation depend on non-integer powers of the energy \citep{Ronco2017,Calcagni2017}.  

Regardless of the model to be used, the expected scale of QG effects is typically several orders of magnitude higher than the energy of observed photons. For this reason we can treat the anomaly induced by QG as a small correction to the photon group velocity and, in particular, only linear $n=1$ or quadratic $n=2$ modifications are of interest for experimental searches taking into account the sensitivity of current detectors. It is important to stress that there are counterexamples where $E_{QG}$ can be far away from the Planck scale (being either above or below $E_P$). Among others, let us highlight two particular cases. In the approach of Asymptotic Safety, renormalization group techniques generate a running of the gravitational constant thereby affecting the value of $E_{QG}$ \citep{Reuter2006}. In String Theory, the compactification of extra dimensions can produce testable effects at energies much lower than $E_P$, even of the order of tens of teraelectronvolts \citep[TeV, ][]{Arkani1998}. Some stringent constraints already exist on these models \citep{ATLAS2016}. Given that, different types of experiments play a crucial role in constraining the value of $E_{QG}$.

To compensate for the smallness of the effect ($E/E_{QG}$ is typically of the order $10^{-19} - 10^{-14}$), it has been recognized that very distant astrophysical sources could be used to probe properties of quantum space-time \citep{Amelino1998, Urrutia1999, Liberati2006}. Indeed, if the emitted photons travel over large distances, then even extremely tiny quantum-space-time effects could accumulate and eventually the overall effect could become detectable in the form of energy-dependent time delays in the light curves. Variable or transient sources at cosmological distances such as GRB and flaring active galactic nuclei (AGN) are good candidates looking for LIV, but it is important to stress that the involvement of cosmological distances forces us to face the problem of combining curvature with quantum-space-time effects. In other words, the delays should depend on the redshift. On the other hand, fast-spinning pulsars (PSRs) detected at TeV energies are within our Galaxy and, thus, their euclidean distances can be used instead of the redshift.

\begin{table*}[t!]
\begin{center}
\caption{A selection of limits for a sub-luminal propagation obtained with various instruments and various types of objects.
\label{tab:all_res}}
\scriptsize
\begin{tabular}{llllllll}
\hline
\hline
Source & Experiment & Year & Distance$^a$ & Lower limit on $E_{QG,1}$ & Lower limit on $E_{QG,2}$ & Reference & Note \\
       &            &              &              & (95\% CL, GeV)            & (95\% CL, GeV)                &          &       \\
\hline
35 GRB & BATSE, HETE-2, Swift & - & - &  $1.4\times10^{16}$ & - & 1 & $^{b}$ \\
8 GRB & Fermi LAT  & - & - & $1.0\times10^{17}$ & - & 2 & \\
\object{GRB 090510} & Fermi LAT & 2009 & 0.903 & $9.3\times10^{19}$ & $1.3\times10^{11}$ & 3 & \\
\object{GRB 190114C} & MAGIC & 2019 & 0.4245 & $0.6\times10^{19}$ & $6.3\times10^{10}$ & 4 &  \\
\object{Mrk~501} & MAGIC & 2005 & 0.034 & $0.3\times10^{18}$ & $5.7\times10^{10}$ & 5   & $^{c,\star}$ \\
\object{Mrk~501} & H.E.S.S. & 2014 & 0.034 & $3.6\times10^{17}$ & $8.5\times10^{10}$ & 6 &\\
\object{PKS 2155-304} & H.E.S.S. & 2006 & 0.116 & $2.1\times10^{18}$ & $6.4\times10^{10}$ &  7 & $^\star$ \\
\object{PG 1553+113} & H.E.S.S. & 2012 & $0.49\pm0.04$ & $4.1\times10^{17}$ & $2.1\times10^{10}$ & 8 & $^{d,\star}$\\
\object{PSR B0531+21} & VERITAS & 2007-14 & 2.2 kpc & $1.9\times10^{17}$ & - & 9 &  \\
\object{PSR B0531+21} & MAGIC   & - & 2.2 kpc & $5.5\times10^{17}$ & $5.9\times10^{10}$ & 10 & $^\star$ \\
\object{PSR B0833-45} & H.E.S.S. & - & 294 pc & $4.0\times10^{15}$ & - & 11 & $^\star$ \\
\hline
\end{tabular}
\end{center}

{Notes.}\\
$^a$ Redshift is given for extra galactic objects.
$^{b}$ The limits of \citet{Ellis2006} were corrected in \citet{Ellis2008} taking into account the factor $(1+z')$ in the numerator of integral in Eq. \protect{\ref{eq:kappaliv}}. Only the limit obtained for a linear correction is given.
$^{c}$ These numbers are actually reported as best fit values by \citet{Martinez2009}.
$^{d}$ The redshift of this source was not measured but only estimated.
$^\star$ Sources used as benchmark in the present paper.\\
{References.}\\
(1) \citet{Ellis2006, Ellis2008}, (2) \citet{Ellis2019}, (3) \citet{Vasileiou2013}, (4) \citet{Acciari2020}, (5) \citet{Martinez2009}, (6) \citet{Abdalla2019} , (7) \citet{Abramowski2011}, (8) \citet{Abramowski2015}, 
(9) \citet{Zitzer2013}, 
(10) \citet{Ahnen2017}, (11) \cite{Chretien2015}.
\end{table*}

Considering only the leading dominant term in Equation~(\ref{eq:disprel1}), either linear ($n=1$) or quadratic ($n=2$), it can be shown that the group velocity of photons acquires a dependence on their energies. In particular, the delay between two photons emitted at the same time by a source at redshift $z$ with energies $E_h > E_l$ is:
\begin{equation}
\label{eq:timez5}
\Delta t_n \simeq \pm\,\frac{n+1}{2}\,\frac{E_h^n - E_l^n}{\mathrm{H}_\mathrm{0} E_{QG}^n}\ \kappa_n(z),
\end{equation}
where $\kappa_n(z)$ is a parameter depending on the distance of the source. The symbol $\pm$ allows to take into account both a subluminal (sign $+$) or a superluminal (sign $-$) LIV effect. In this paper, two different expressions for $\kappa_n(z)$ will be compared for the first time: one obtained in a pure Lorentz invariance violation framework \citep{Jacob2008}, and another obtained in the doubly special relativity (DSR) approach \citep{Rosati2015}. This will be discussed in more detail in Section~\ref{sec:dist}.

The delay $\Delta t_n$ takes only into account Lorentz violation effects, therefore neglecting any time lag originating from emission mechanisms, also referred to as `source intrinsic' delay. A hint of such kind of delay are observed for GRB \citep{Ajello2019} and has been also reported once in the case of an AGN, for the flare of \object{Mrk~501} in 2005 recorded by MAGIC\,\footnote{\textit{Major Atmospheric Gamma Imaging Cherenkov}, \url{https://magic.mpp.mpg.de}} \citep{Albert2007}. With only one source, and with only a rough knowledge of how particles are emitted and accelerated, intrinsic delays cannot be separated from propagation effects. Modeling of astrophysical sources is an on-going effort and a first study of source intrinsic effects in connection with Lorentz violation searches has been published recently in the case of blazar flares \citep{Perennes2020, Levy2021}. On the other hand, when several sources are combined, it could be possible, at least in principle, to separate intrinsic and propagation effects. Indeed, it is reasonable to assume that intrinsic delays do not depend on the distance. It is therefore essential that these studies could be performed on a large population of objects. 

From Equation~(\ref{eq:timez5}), another parameter $\lambda_n$ can be defined as
\begin{equation}
\label{eq:lambda}
\lambda_n \equiv \frac{\Delta t_n}{\Delta E_n\ \kappa_n(z)} = \pm \frac{n+1}{2 \mathrm{H}_\mathrm{0}\ E^{n}_{QG}},
\end{equation}
using the simplified notation $\Delta E_n \equiv E_h^n - E_l^n$. This parameter $\lambda_n$, which will be used later, has the advantage to be independent of the distance of the source and is therefore suitable for a multi-source analysis.

Since the late 90s, the field has rapidly expanded, with more and more sources being analyzed in the search for Lorentz invariance violation effects. With the notable exception of the flare of \object{Mrk~501} in 2005 already mentioned above, no significant delay has been reported so far when using only photon as the messenger. Constraints have been improving on a regular basis, even reaching the Planck scale in some cases in analyses of individual objects \citep[see e.g.][]{Vasileiou2013}. \mbox{Table}~\ref{tab:all_res} gives a partial selection of the best limits available on $E_{QG,n}$ for time-of-flight studies, with the three types of sources (AGN, GRB and PSR). This new notation $E_{QG,n}$ reflects the fact that LIV analyses have different sensitivities for linear and quadratic effects.

 The results of \cite{Ellis2006, Ellis2008, Ellis2019} are of particular interest since they were obtained from the analyses of several GRB. This kind of analysis, repeated with different experiments \citep[e.g.][]{Bolmont2008, Bernardini2017} consists in two steps: first, the time-lags are computed for each individual source, and then the obtained data points are fitted with a function $\Delta t = a\,z + b\,(1 + z)$ (for $n=1$). The value of parameter $a$ is subsequently used to constrain $E_{QG,1}$ while $b$ represents source intrinsic effects, assumed to be identical for all bursts. In the present paper, we describe and test a more advanced and presumably more sensitive method to perform such a population study, based on a likelihood technique. 

For completeness, let us mention that recently it was suggested from the analysis of several GRB that Lorentz invariance could be violated at a scale of \mbox{$\sim3.7\times 10^{17}$~GeV} \citep{XuMa2016a, XuMa2016b, XuMa2018}. This result is contradictory with the best limits listed in Table~\ref{tab:all_res} and still needs to be confirmed.

One of the main objectives of the first phase of describing QG phenomenology was the ability to prove that in-vacuo dispersion (or, more generally, Planck-scale effects) could be tested with current experiments. Now, the stringent limits established with GRB observations together with the growing amount of relevant data, and progress on the theory side, can bring us to a more mature phase where we can start constraining actual QG models, in a robust manner.  The present work can be considered a first step in this direction. We aim to combine, for the first time, the data obtained with the three major imaging atmospheric Cherenkov telescope (IACT) experiments, H.E.S.S.\,\footnote{\textit{High Energy Stereoscopic System}, \url{https://www.mpi-hd.mpg.de/hfm/HESS/}}, MAGIC and VERITAS\,\footnote{\textit{Very Energetic Radiation Imaging Telescope Array System}, \url{https://veritas.sao.arizona.edu}} in order to constrain QG effects through the time-of-flight technique \citep[see][for a recent review]{terzic2021}. This combination will extract the most information out of each type of source to produce robust constraints from existing data, while taking into account the redshift dependence of the LIV-induced time-lag. \\


The paper will be divided in two parts. In the present article (part I), two possible ways to account for the dependence of time delays as a function of redshift will first be described (Section~\ref{sec:dist}). Then, the method used to compute and combine the likelihoods to measure time-lag parameters $\lambda_n$ and $\tau_n$ will be described in detail (Section~\ref{sec:method}). Several nuisance parameters are included in the computation to take into account various sources of systematic uncertainties. Then, in Section~\ref{sec:simulations}, the method is tested on simulated data sets mimicking the data of several representative sources observed in the TeV domain. These simulations are used to evaluate statistical errors and study the impact of various sources of systematic errors in the lag measurement. The results, as well as the impact of redshift dependence, will be given and discussed in Section~\ref{sec:res}.

In the second part of the paper, to appear later, the method will be used with available data from H.E.S.S., MAGIC and VERITAS, and possibly from other gamma-ray experiments, in order to produce a combined limit on $E_{QG,n}$.

\section{Redshift dependence of time delays}\label{sec:dist}

It is rather natural to believe that curvature and quantum effects are deeply intertwined since curvature is a key characteristic of space-time geometry. In light of this, a complete QG theory would be needed to tell us whether there is a phenomenon of in-vacuo dispersion and then compute its magnitude. However, in the absence of such a theory, simplified speculative approaches to model in-vacuo dispersion in curved spaces have been proposed. Among them, especially for reasons of simplicity, a model where Lorentz invariance is explicitly broken in a specific way proposed by Jacob and Piran \citep[][J\&P for short]{Jacob2008} attracted a particular interest and has been systematically used so far in experimental analyses constraining in-vacuo dispersion. 

\begin{figure}
    \plotone{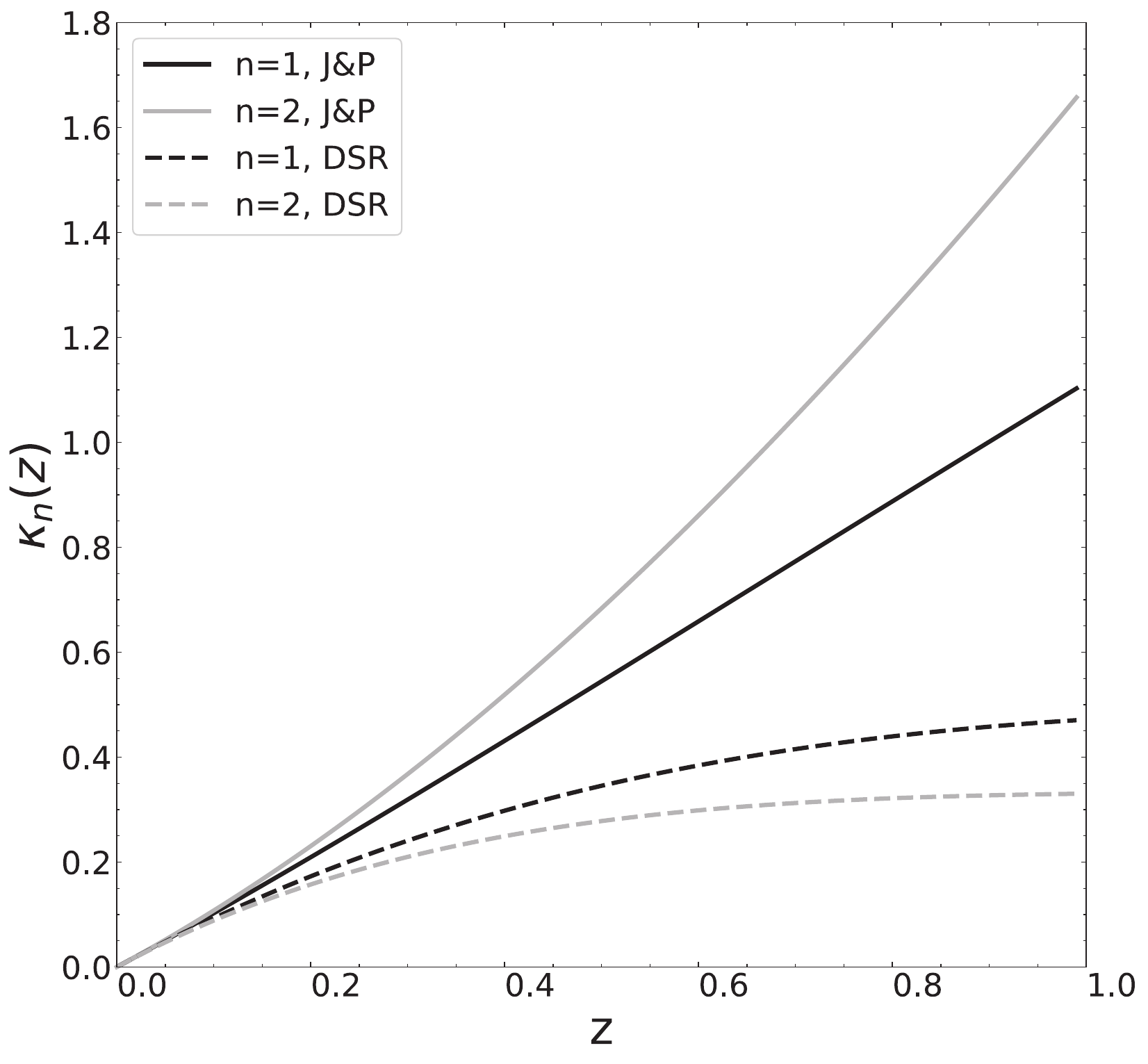}
    \caption{Parameter $\kappa$ for $n=1$ (black) and $n=2$ (gray) in the J\&P case (solid line) and in the DSR case (dashed line).}
    \label{fig:kappaz}
\end{figure}

In this approach, parameter $\kappa_n(z)$ is expressed as:
\begin{equation}
\label{eq:kappaliv}
\kappa^\mathrm{J\&P}_n(z) \equiv  \int_0^z \frac{(1+z')^n}{\sqrt{\Omega_m\,(1+z')^3 + \Omega_\Lambda}}\ dz',
\end{equation}
where {the denominator relates to the Hubble parameter} $H(z) = \mathrm{H}_\mathrm{0} \sqrt{\Omega_m\,(1+z)^3 + \Omega_\Lambda}$. In the following, cosmological parameters values are taken from Planck results \citep{Planck:2018vyg} as recommended by the Particle Data Group \citep{Zyla:2020zbs}: $\mathrm{H}_\mathrm{0} = 67.4\pm0.5\ \mathrm{km\,s}^{-1}\,\mathrm{Mpc}^{-1}$, $\Omega_m = 0.315\pm0.007$ and $\Omega_\Lambda = 0.685\pm0.007$.

As recent literature has pointed out \citep{Rosati2015,Barcaroli2016,Pfeifer2018}, Equation~(\ref{eq:kappaliv}) offers only one possible parameterization among many others. It has been shown that if, following the Deformed Special Relativity (DSR) approach, Poincar\'e symmetries are modified in order to preserve the invariance of Equation~(\ref{eq:disprel1}) under relativistic transformations, then one can obtain a different result for the distance parameter \citep{Rosati2015}: 
\begin{equation}
\label{eq:kappadsr}
\kappa^\mathrm{DSR}_n(z) \equiv  \int_0^z \frac{h ^{2n}(z') dz'}{(1+z')^n\, \sqrt{\Omega_m\,(1+z')^3 + \Omega_\Lambda}},
\end{equation}
with
\begin{equation}
    \begin{split}
    h (z') \equiv  1+ z' - &\sqrt{\Omega_m\,(1+z')^3 + \Omega_\Lambda}\\
    & \times \int_0^{z'} \frac{dz''}{\sqrt{\Omega_m\,(1+z'')^3 + \Omega_\Lambda}}\,
    \end{split}
\end{equation}
and this can result in consistently different limits on $E_{QG,n}$. Note that Equation~(\ref{eq:kappadsr}) is only one possible outcome of DSR, chosen here as a benchmark. Using observations of multiple sources at different redshifts, we establish for the first time limits on these two different models.

Figure~\ref{fig:kappaz} shows functions $\kappa^\mathrm{DSR}$ and $\kappa^\mathrm{J\&P}$ as a function of redshift for $n=1$ and $n=2$. $\kappa^\mathrm{DSR}$ is smaller than $\kappa^\mathrm{J\&P}$ for both linear and quadratic cases. When no lag is measured, this leads to less stringent limits on $E_{QG,n}$. 
{Both functions $\kappa^\mathrm{J\&P}$ and $\kappa^\mathrm{DSR}$ increase with redshift, thus increasing the expected time delay. However, it has to be pointed out that ultimately, the distance at which sources can be detected at high energies is limited by the absorption by the extragalactic background light (EBL). This distance depends on the energy, and does not exceed $z\sim1$ in the TeV range.}

To conclude this section, let us add that in case of nearby sources, such as pulsars, the euclidean approximation is valid, i.e. $\kappa_n(z) = d\,\mathrm{H}_\mathrm{0}/c$ where $d$ is the euclidean distance to the source. In addition, the ratio $\kappa^\mathrm{DSR}/\kappa^\mathrm{J\&P}$ converges to unity for low distances. As a result, a given pulsar will give the same constraints on $E_{QG,n}$ for both J\&P and DSR cases.

\section{Methodology}\label{sec:method}

All observations considered for combination in this work are analyzed with a Maximum Likelihood (ML) method to search for linear or quadratic LIV delays and to extract limits on $E_{QG,n}$. Compared to alternative methods such as \textit{e.g.} PairView \citep{Vasileiou2013}, Dispersion Correction \citep{Barres2012}, Peak comparison \citep{Ahnen2017}, the ML method allows optimal use of the information in data and provides a relatively straightforward way to combine analyses of multiple sources and observatories. On the other hand, it relies on parameterization of intrinsic photon emission time and energy distributions, which are currently not fully understood at the theoretical level. The uncertainties related to these parameterizations are taken into account when deriving the limits on $E_{QG,n}$.

The code for likelihood computation as well as for simulations was developed using the ROOT\,\footnote{\url{https://root.cern.ch}} framework \citep{ROOT}.

\subsection{Single Source Likelihood}

First applied by~\cite{Martinez2009} for analyzing the 2005 flare of \object{Mrk~501} observed by MAGIC, the ML method relies on defining a probability density function (PDF) that describes the probability of observing a gamma-ray photon at a certain arrival time and with a certain reconstructed energy.

In its simplest form, the PDF for signal photons is defined as a function of time and energy with $\lambda_n$ as the single parameter to be estimated. The PDF is obtained convolving the spectrum of the source $\Gamma_{s}(E_t)$ with the light curve $C_s$, both as observed on Earth, \textit{i.e.} after propagation:
\begin{equation}\label{eq:pdf source}
F_{s}(E_{t},t;\lambda_n) = \frac{\Gamma_{s}(E_t)\,C_{s}\left(t-D(E_{t},\lambda_n,z)\right)}{N_s},
\end{equation}
 where $E_{t}$ is the true energy of the gamma-ray photon, $t$ is the arrival time corrected by the factor:
 \begin{equation}\label{eq:Delay}
 D(E_{t},\lambda_n,z) = \lambda_n \times \kappa_n(z) \times E_t^n,
 \end{equation}
 which defines the propagation delay due to LIV, where $\lambda_n$ is given by Equation~(\ref{eq:lambda}), and $N_s$ is a normalization term expressed as follows:
 \begin{equation}\label{eq:pdf sourcenorm}
N_s = \iint \Gamma_{s}(E_t)\,C_{s}\left(t-D(E_{t},\lambda_n,z)\right)\,dE_t\,dt.
\end{equation}

In Equations~(\ref{eq:pdf source}) and (\ref{eq:pdf sourcenorm}), function $C_s$ is often called the \textit{template} light curve. It is usually obtained by fitting a light curve at low energies, where LIV effects are assumed to be weak or negligible. Since there is no fully accepted model available which reproduces the shape of the light curves for GRB, AGN or PSR, Gaussian or Lorentzian functions, or the sum of several of these functions are usually used. {The function $\Gamma_{s}$ is obtained from the data on the full energy range considered for the LIV analysis (see Section~\ref{sec:simulations}).}

\begin{table*}[t!]
      \caption{Nuisance parameter uncertainties for the individual sources.}\label{tab:nuisance}
\begin{center}
\scriptsize
        \begin{tabular}{llllll} 
        \hline
        \hline
        Source & Energy Scale & Background proportion & Spectral index$^a$ & Distance/redshift & References$^b$ \\
        
        \hline
        GRB 190114C & $17$\% & $11$\% & $0.21$ & $\Delta z = 1\times10^{-3}$ & 1, 2, 3, 4\\
        PG 1553+113 & $10$\% & $20$\% & $0.31$ & $\Delta z = 4\times10^{-2}$ & 5, 6\\ 
        Mrk~501 & $17$\% & $11$\% & $0.04$ & $\Delta z = 1\times10^{-4}$ & 1, 2, 3, 6\\ 
        PKS 2155-304 & $10$\% & $20$\% & $0.1$ & $\Delta z = 1.7\times10^{-2}$ & 5, 7\\ 
        Crab (M) & $17$\% &  $11$\%  & $0.07$ & $\Delta d = 506$ pc & 1, 2, 3, 8\\
        Crab (V) & $20$\% & $22$\% & $0.5$ & $\Delta d = 506$ pc & 9, 8\\
        Vela & $10$\% & $20$\% & $0.67$ & $\Delta d = 76$ pc & 5, 10\\ \hline
        \end{tabular}
  
\end{center}
{Note.}\\
$^a$ Uncertainty for spectral index includes both statistical and systematic errors. $^b$ References are in the same order as the columns: energy scale, background proportion, spectral index (the three values are sometimes given in the same reference), and distance.\\
{References.}\\
(1) \citet{Aleksic2012},
(2) \citet{Aleksic2016},
(3) \citet{Aleksic2015},
(4) \citet{Acciari2019b},
(5) \citet{Aharonian2006},
(6) \citet{Mao2011},
(7) \citet{Ganguly2013},
(8) \citet{Kaplan2008},
(9) For energy scale and spectral index, \citet{Pueschel2019}. The value for background proportion was provided by the VERITAS collaboration,
(10) \citet{Caraveo2001}
\end{table*}

Background events of several different origins are also taken into account. They include hadrons mis-reconstructed as gamma-rays and baseline photons emitted either by an AGN in its quiescent state or by the nebula surrounding a PSR. The PDF for background events of type \textit{k} (hadrons or baseline photons), which are not affected by LIV propagation effects, is written as:
\begin{equation}\label{eq:pdf bkg}
F_{b,k}(E_{t},t) = \frac{\Gamma_{b,k}(E_t)\,C_{b,k}}{N_{b,k}}.
\end{equation}
$\Gamma_{b,k}$ is the background spectrum taken as a power law. For hadrons, the index is set to 2.7 while the values for signal and baseline photons are given in Table~\ref{tab:simulation}. $C_{b,k}$ is the time distribution of background events assumed to be a constant, and $N_{b,k}$ the normalization term defined as:
 \begin{equation}\label{eq:pdf bkgnorm}
N_{b,k} = \iint \Gamma_{b,k}(E_t)\,C_{b,k}\,dE_t dt.
\end{equation}

From Equations~(\ref{eq:pdf source}) and (\ref{eq:pdf bkg}), the complete definition of the PDF is obtained, accounting for detector performance assessed from instrument response functions (IRFs):
\begin{multline} \label{eq:pdf source+det}
\frac{dP}{dE_m dt}= w_s\ \frac{\int A(E_{t}, \vec{\varepsilon}) M(E_t,E_m) \times F_{s}(E_{t},t;\lambda_n) dE_{t}}{N'_{s}} \\
+ \sum_{k} w_{b,k}\ \frac{\int A(E_{t}, \vec{\varepsilon}) M(E_t,E_m) \times F_{b,k}(E_{t},t) dE_{t}}{N'_{b,k}},
\end{multline}
where the source and background terms $F_s$ and $F_{b,k}$ are convoluted  with the detector effective area $A(E_{t},\vec{\varepsilon})$ and energy resolution $M(E_t,E_m)$. Source and background terms are weighted by $w_s$ and $w_{b,k}$ respectively, with $w_s+\sum_{k} w_{b,k} = 1$. $E_t$ still denotes the true energy while $E_m$ is the corresponding measured energy. Parameters $N'_{s}$ and $N'_{b,k}$ are the normalization factors of the PDF. In addition to energy $E_t$, effective area depends on a set of factors $\vec{\varepsilon}$ which vary with observation conditions and with the method used for event reconstruction and identification. The IRFs were kindly provided by the H.E.S.S., MAGIC and VERITAS collaborations. Distinct IRFs are used for each source and each observation period.

The confidence levels for either a measurement or the derivation of lower limits on $\lambda_n$ (and $E_{QG,n}$) can then be obtained summing the log-likelihood of all the events for a given source $S$:
\begin{equation}\label{eq:LikelihoodData}
L_{S}(\lambda_n) = -\sum_{\mathrm{i}} \log\left(\frac{dP}{dE_m dt}(E_{m,i},t_{i});\lambda_n\right).
\end{equation}

\subsection{Combining Likelihoods}

While each source may require a different analysis strategy, either using a single parameter likelihood or a profile likelihood, the combination of multiple sources is straightforward. Once log-likelihood functions $L_{S}(\lambda_n)$ are obtained for all sources, the combined log-likelihood $L_{comb}$ is simply given by their sum:
\begin{equation}\label{eq:combination of likelihood}
L_{comb}(\lambda_n) = \sum_{\mathrm{all\ sources}} L_{S}(\lambda_n).
\end{equation}

\subsection{Statistical and systematic uncertainties}\label{subsec:uncertainties}

Statistical and systematic uncertainties are propagated in the final result through the use of profile likelihood. The log-likelihood for each source is then written as:
\begin{multline}
L(\lambda_n,\vec{\theta}) = L_\mathrm{S}(\lambda_n,\vec{\theta}) + L_\mathrm{template}(\vec{\theta}_\mathrm{C}) + L_\mathrm{\gamma}(\theta_\mathrm{\gamma}) +\\ 
L_\mathrm{B}(\vec{\theta}_\mathrm{B}) + L_\mathrm{ES}(\theta_\mathrm{ES}) + L_\mathrm{z}(\theta_\mathrm{z}),
\end{multline}
where $\vec{\theta}$ is the vector of all nuisance parameters defined as: 
\begin{enumerate}[label=\alph*.]
\item $\vec{\theta}_\mathrm{C}$, the parameters of the light curve analytic parameterization,
\item $\theta_\mathrm{\gamma}$, the power law index of signal events spectrum,
\item $\vec{\theta}_\mathrm{B}$, the ratio of signal and of background event numbers to the total number of events,
\item $\theta_\mathrm{ES}$, the energy scale,
\item $\theta_\mathrm{z}$, the distance or redshift.
\end{enumerate}

\begin{table*}[t!]
      \caption{Simulation settings for the individual sources. \label{tab:simulation}}
\begin{center}
\scriptsize
        \begin{tabular}{lllllll} 
        \hline
        \hline
        Source & Energy Range & Time Range$^a$ & Spectral index & Lightcurve shape & Number of events  & Background proportion  \\
         & (TeV) & & $\Gamma_s$, $\Gamma_b$ & & likelihood$^b$, template$^c$ & hadronic, baseline \\
        
        \hline
        GRB 190114C & $0.3$ - $2$ & $60$ - $1200$ s & $5.43$, - & curved power law & $726$, - & $0.055$, $0.$\\
        PG 1553+113 & $0.4$ - $0.8$ & $0$ - $8000$ s & $4.8$, $4.8$ & double Gauss  & $72$, $82$ & $0.29$, $0.15$ \\ 
        Mrk~501 & $0.25$ - $11$ & 0 - $1531$ s & $2.2$, $2.2$ & single Gauss & $1800$, - & $0.39$, $0.$\\ 
        PKS 2155-304 & $0.28$ - $4$ & $0$ - $4000$ s & $3.46$, $3.32$ & 5 asymmetric Gauss & $2965$, $561$ & $0.$, $0.02$ \\ 
        Crab (M) & $0.4$ - $7$ & $0.36$ - $0.45$ & $2.81$, $2.47$ & single Gauss + Baseline & $14869$, - &  $0.$, $0.961$ \\
        Crab (V) & $0.2$ - $10$ & $0.37$ - $0.43$ & $3.25$, $2.47$ & single Gauss + Baseline & $22764$, - &  $0.$, $0.964$\\
        Vela & $0.06$ - $0.15$ & $0.50$ - $0.60$ & $3.9$, $1.75$ & asymmetric Lorentzian  & $330820$, - & $0.$, $0.998$ \\ \hline
        \end{tabular}
      
\end{center}
{Notes.}\\
$^a$ For pulsars, the phase range is given, \textit{i.e.} the time range normalized with respect to the rotation period.
$^b$ Number of photons considered when computing the likelihood, \textit{i.e.} excluding the ones used for template determination. $^c$ A sign '-' means no template was used (see Section~\ref{subsec:uncertainties} for details).
\end{table*}

As already mentioned above, the template light curve $C_s$ of Equation~(\ref{eq:pdf source}) is obtained by fitting a low energy light curve, for which LIV is assumed to be negligible. From this parameterization, it is possible to evaluate errors directly, defining $L_\mathrm{template}(\vec{\theta}_\mathrm{C})$ as the sum of the log-likelihoods of each event generated from the low energy template parameterization:
\begin{equation}
L_\mathrm{template}(\vec{\theta}_\mathrm{C}) = -\sum^{N_{\mathrm{template}}}_{i=1} \log\left( \frac{C_s(t_i,\vec{\theta}_\mathrm{C})}{N_c}\right),
\end{equation}
with $C_s$ the light curve and $N_c$ its normalization. In this equation, the new notation for the light curve $C_s(t_i,\vec{\theta}_\mathrm{C})$ denotes the fact the template is evaluated for a zero-lag ($D(E_{i},\lambda_n = 0,z) = 0$), and explicitly shows the parameter vector $\vec{\theta}_\mathrm{C}$ of the template function. On the other hand, some other analyses use the template fit results as nuisance parameters. In that case, $L_\mathrm{template}(\vec{\theta}_\mathrm{C}) = 0$ and the uncertainty on $\vec{\theta}_\mathrm{C}$ is then accounted for in the generated data sample log-likelihood $L_\mathrm{S}(\lambda,\vec{\theta})$ defined in Equation~(\ref{eq:LikelihoodData}).

$L_\mathrm{\gamma}(\theta_\mathrm{\gamma})$ is obtained from the statistical and systematical uncertainties of the spectral index as provided in the analyses of the different sources. The flux normalization and energy scale uncertainties provided by the different observatories are taken into account by  $L_\mathrm{B}(\vec{\theta}_\mathrm{B})$ and  $L_\mathrm{ES}(\theta_\mathrm{ES})$, respectively. The energy scale parameter is introduced in the data sample log-likelihood by a scale factor applied to the event energy. The uncertainties on redshift for extragalactic sources, or distance for galactic sources, are accounted for in $L_\mathrm{z}(\theta_\mathrm{z})$.

For the power law index, ratio of signal and of background, energy scale and redshift uncertainties, a normal distribution is assumed which allows to use a simple chi-square approach: 
\begin{equation}
L_\mathrm{x}(\vec{\theta}_\mathrm{x}) = \sum_{i} \frac{(\theta_{\mathrm{x},i} - \Bar{\theta}_{\mathrm{x},i})^2}{2\sigma^2_{\theta_{\mathrm{x},i}}},
\end{equation}
where $\sigma^2_\theta$ is the uncertainty of the nuisance parameter $\theta$ and $x$ is the different type of systematics. The full list of uncertainties assigned to each nuisance parameter for each source is shown in Table~\ref{tab:nuisance}.

In order to illustrate the impact of the different sources of uncertainties, the uncertainty on lambda is derived by varying only one of the different nuisance parameters. The systematic errors are then derived assuming the total uncertainty is the squared sum of statistical and systematical uncertainty. They are presented in the Appendix with Table~\ref{tab:systematics_res} for the J\&P case and Table~\ref{tab:systematics_res_DSR} for the DSR case, for each source and each source combination. These results will be commented further in Section~\ref{sec:res}.

\section{Simulations}\label{sec:simulations}

\subsection{Simulated data sets}\label{subsec:sim_data sets}

\subsubsection{Data sets choice criteria}\label{subsubsec:srcs_criteria}

The sources used in this study are listed in Table~\ref{tab:simulation}. They have all been detected by the three experiments H.E.S.S., MAGIC and VERITAS, and have been selected to gather a representative sample. Namely, the three types of source were selected: one GRB, three flaring AGN and two PSR, with LIV results already published, and with the following additional criteria:
\begin{enumerate}[label=\alph*.]
    \item The three flaring AGN show different signal to background ratios: negligible background for \object{PKS~2155-304} and \object{Mrk~501} and substantial background level for \object{PG~1553+113},
    \item The sources show very different light curve shapes, from a  single Gaussian pulse for \object{Mrk~501} flare of 2005 to multiple asymmetric spikes for \object{PKS~2155-304} flare of 2006,
    \item The sources selected cover a wide range in distance, from 2~kpc for the Crab PSR to a redshift of 0.49 for \object{PG~1553+113},
    \item The two PSR have different distances and were observed on very different time scales,
    \item In addition, \object{PG~1553+113} has a large uncertainty on the distance which was taken into account in the analysis.
\end{enumerate}

In the following sub-section, the most important characteristics of the sources as taken from the references listed in Table~\ref{tab:all_res} are briefly summarized. The numbers given in Table~\ref{tab:simulation} were extracted from these references or provided by the authors in private communications. Then, the use of simulated data-sets to assess the performance of the method is described in Section~\ref{subsec:calperf}.

Except specified otherwise, a spectral index $\Gamma_k = 2.7$ was used for hadrons.

\subsubsection{Source description}\label{subsubsec:srcs}

\object{GRB~190114C} is a gamma-ray burst detected on 2019 January 14 at 20:57:03 Universal Time (UT) and located at redshift $z=0.4245 \pm 0.005$~\citep{Selsing2019,Castro-Tirado2019}. Following the alert sent by \textit{Swift} \citep{Gropp2019}, MAGIC observed \object{GRB~190114C}, detecting a strong VHE $\gamma$-ray signal \citep{Acciari2019a}. The observations started 62 seconds after the beginning of the burst. A total of $\sim$700 events with energy ranging from 300~GeV to $\sim$2~TeV were recorded during the first 19 minutes of observations. The intrinsic energy distribution of the signal was fitted with a power law of index $2.5 \pm 0.2$, leading to an index of $\Gamma_s = 5.43\pm 0.22$ (statistical error only) when EBL absorption is taken into account. The time distribution of the events recorded by MAGIC follows a power law with index $1.51 \pm 0.04$. MAGIC did not observe the peak of the burst. Therefore, the light curve of the full burst, including the sharp rise to the peak flux followed by a power-law decay was modeled based on multiwavelength observations of the event and theoretical inference~\citep{Acciari2019b}. 
{The prompt emission of GRB~190114C inferred from the keV--MeV light curves and spectra lasted no more than $25$ seconds after the onset of the GRB. This indicates that the emission observed by MAGIC is associated with the afterglow phase, rather than with the prompt phase, which typically shows irregular variability. However, as reported in \citet{Acciari2020}, a sub-dominant contribution from the prompt phase (at most 20\%) at early times of the afterglow ($t \lesssim 100$~s) cannot be entirely excluded. The lower bound of 60~s chosen for the present study was chosen to minimize this contribution while retaining statistics as high as possible.}

\object{Mrk~501} is a BL Lac object at redshift $z=0.03364$. The flare of 2005 July 9 was detected by the MAGIC telescope, at the time operating in monoscopic configuration \citep{Albert2007}. The flux of this flare reached a peak more than a factor of two higher than before and after the flare. A total of $\sim$1800 events with energy from 0.15 to 10 TeV were recorded during the flare among which $\sim$700 could be associated to the background. The energy distribution of the signal and baseline events is well described by a power law of index $\Gamma_{s,b} = 2.2$ while the time distribution was parameterized by a single Gaussian spanning over $1600$ seconds.

\object{PKS 2155-304} is another BL Lac object at higher redshift $z=0.116$. The flare of 2006 July 28 detected by H.E.S.S. telescopes is seemingly one of the brightest flares detected by the experiment so far with a signal to noise ratio exceeding 300 \citep{Aharonian2007}. The lightcurve is parameterized by five asymmetric Gaussians with $2\%$ background over $4000$ seconds with a total of $3526$ photons. The energy distribution is described by a power law of index $\Gamma_s = 3.46$ ranging from 0.25 to 4~TeV during the flare while the quiescent state leads to an index of~$\Gamma_b = 3.32$.

\object{PG 1553+113}, yet another BL Lac object, is the furthest source of this list with an estimated redshift $z=0.49\pm0.04$. The flare of 2012 April 26-27 was detected by H.E.S.S. telescopes where its flux increased three-fold as compared to its quiescent state \citep{Abramowski2015}. The time distribution was parameterized by two Gaussians with 154 photons over $8000$ seconds, where background accounts for $44\%$ of the events with $30\%$ gamma-like hadrons and $14\%$ baseline photons. The energy distribution spreading between $0.3$ and $0.8$ TeV is described by a power law of index $\Gamma_{s,b} = 4.8$ for signal and baseline photons.

The Vela Pulsar (\object{PSR B0833-45}) located at $294\pm76$~pc rotates with a periodicity of $89$ ms. The data simulated in this work is from a compilation of observations with H.E.S.S. large telescope from March 2013 to April 2014, for which a LIV analysis was performed \citep{Chretien2015, ChretienPhD}. {330,820 pulsed events} between 60 and 150~GeV were recorded with a signal to noise ratio of $0.012$. The phase distribution is parameterized by an asymmetric Lorentzian between $0.5$ and $0.6$. Background accounts for $98.8\%$ of the events with only baseline photons. The energy distribution is described by a power law of index $\Gamma_s = 3.9$ for signal and $\Gamma_b = 1.75$ for baseline photons.

\begin{figure}[t!]
    \plotone{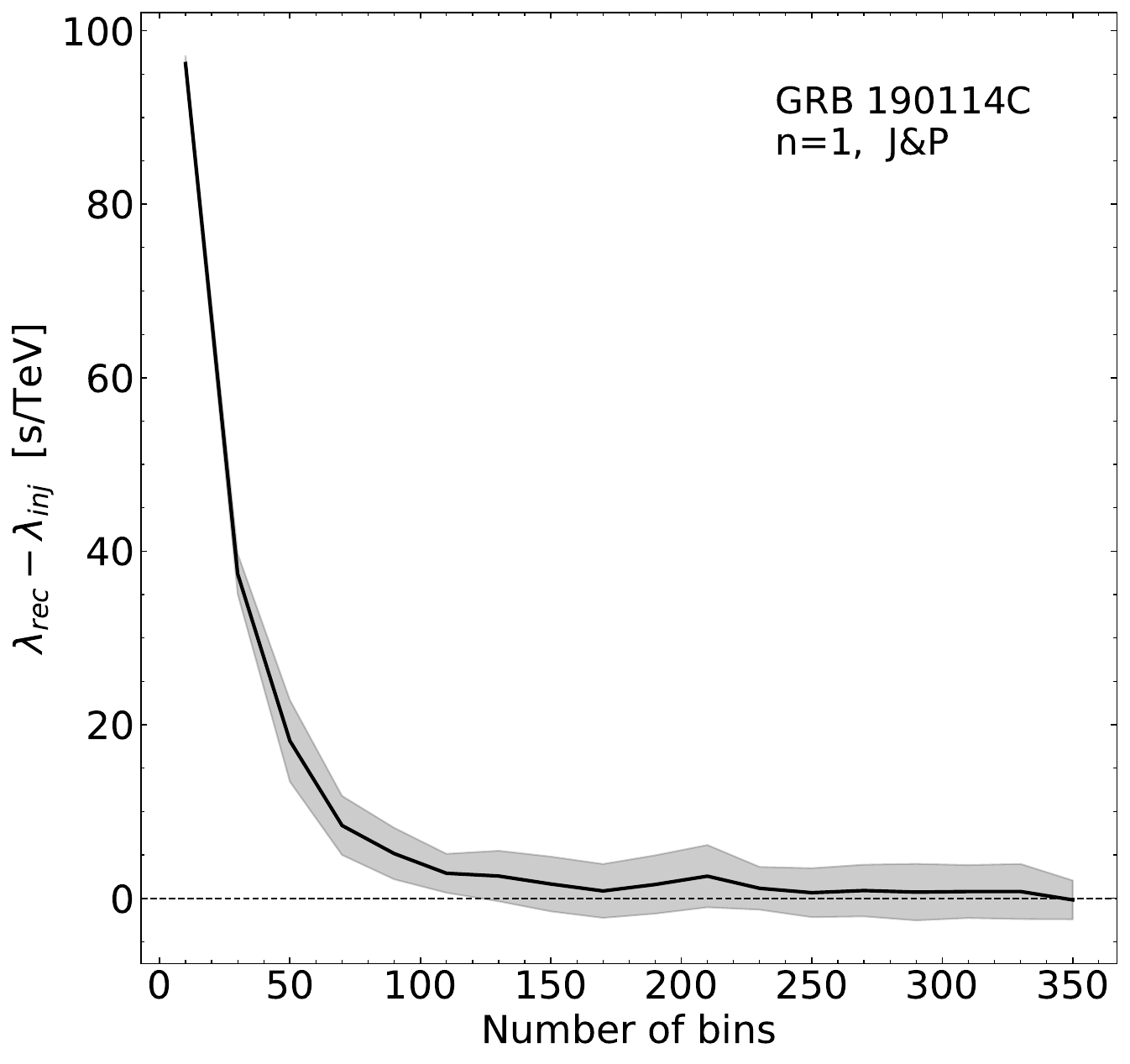}
    \caption{Bias $\lambda_{rec} - \lambda_{inj}$ vs. number of bins for GRB~190114C in the linear case and J\&P formalism. The number of bins in the table is chosen so that the bias (black line) is compatible with zero within its 1$\sigma$ uncertainty range (gray envelope). The same number of bins is used for measured energy, non-delayed arrival times, and time delays.}
    \label{fig:convergence}
\end{figure}

The Crab Pulsar (\object{PSR B0531+21}) has a $33.7$ ms period and is located at $2.0 \pm 0.5$ kpc. One of the data sets used in this work, noted ``Crab M'' hereafter, is a compilation of observations with MAGIC telescopes from 2005 to 2017 \citep{Ahnen2017}. $3080 \pm 460$ excess events from the P2 region of the phase were recorded, from which $544 \pm 92$ have a reconstructed energy above 400~GeV and are used in the LIV analysis. The phase distribution of the P2 peak was parameterized by a Gaussian. A profiling of the nuisance parameters yielded a mean of $\Phi=0.403$ (respectively $0.401$) and standard deviation $0.015$ ($0.011$) for $n=1$ ($n=2$). Background accounts for 96\% of the events with only baseline photons. The energy distribution was described by a power law of index $\Gamma_s = 2.81$ for signal and $\Gamma_{b,k} = 2.47$ for combined background events and baseline photons.

The other data set, noted ``Crab V'', is a compilation of high quality data taken with VERITAS telescopes between 2007 and 2011. {22,764 pulsed events} were recorded from the P2 region and its baseline where background account for $96.4\%$ of the events with again only baseline photons \citep{Zitzer2013}. The phase distribution was also parameterized with a Gaussian centered on $0.398$ with a standard deviation of $0.0116$. The energy distribution was again described by a power law of index $\Gamma_s = 3.25$ for signal and $\Gamma_b = 2.47$ for baseline photons.

\subsection{Method calibration and performance}\label{subsec:calperf}

The normalization factor $N'_{s}$ of the PDF of Equation~(\ref{eq:pdf source+det}) is a triple integral, computation of which is particularly time consuming since it needs to be done for each minimization step and for each event of the sample. To decrease the computation time, the PDF is pre-calculated and stored in tables binned over measured energy $E_{m}$, non-delayed arrival times $t$, and time-delays $D(E_t,\lambda_n,z)$. The same number of bins is used for each of these three variables. A {trilinear} interpolation is performed on these tables to extract PDF values for the likelihood computation.

\begin{figure*}[t]
    \plotone{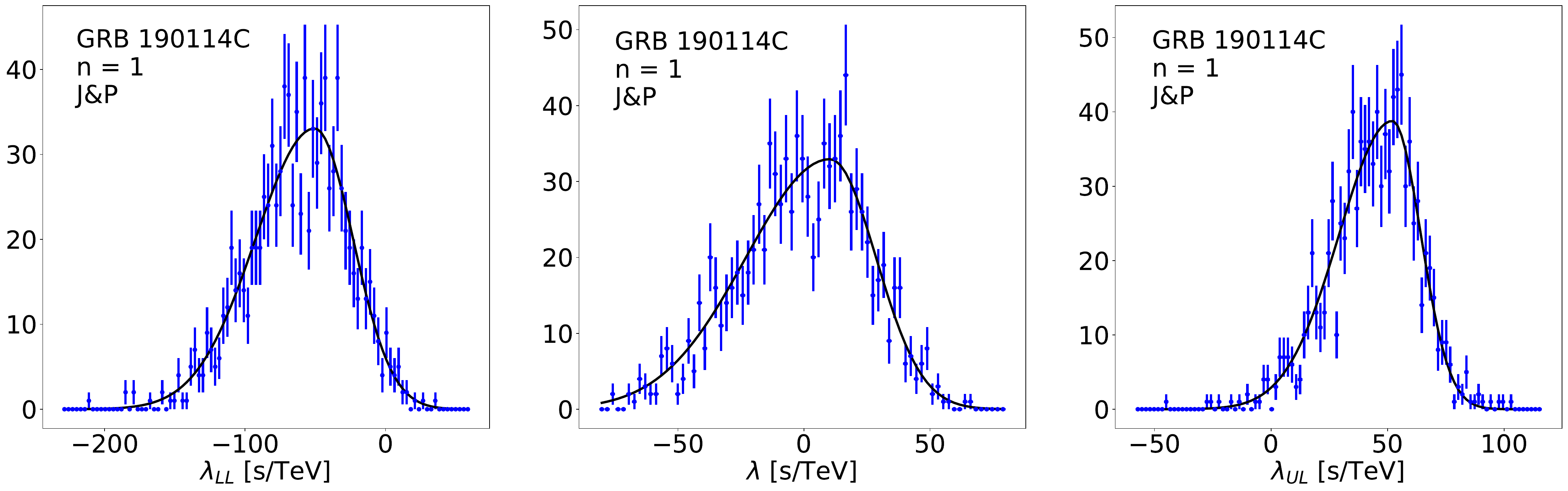}
    \caption{The center plot shows the distribution of the reconstructed lag in the case of GRB~190114C, J\&P formalism for the linear case. The plot on the left (respectively on the right) shows the distribution of the lower (upper) limits of the confidence interval for 68\% CL. The three distributions are obtained with a zero injected lag. The histograms are fitted with asymmetric Gaussian functions, parameters of which are used in turn to produce calibration plots (see the text for details).}
    \label{fig:LL_UL_distribs}
\end{figure*}

\begin{figure*}[t!]
    \plottwo{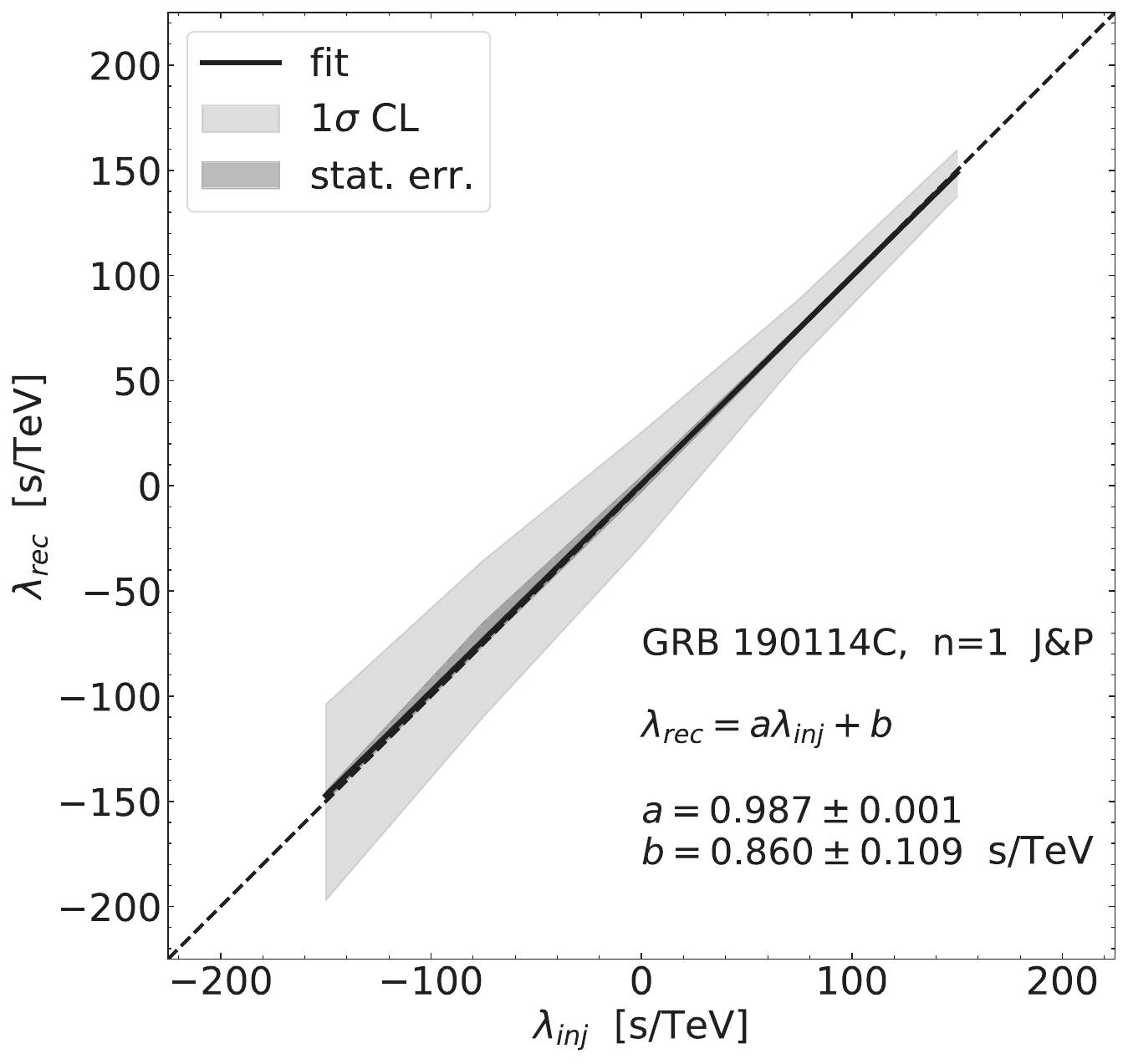}{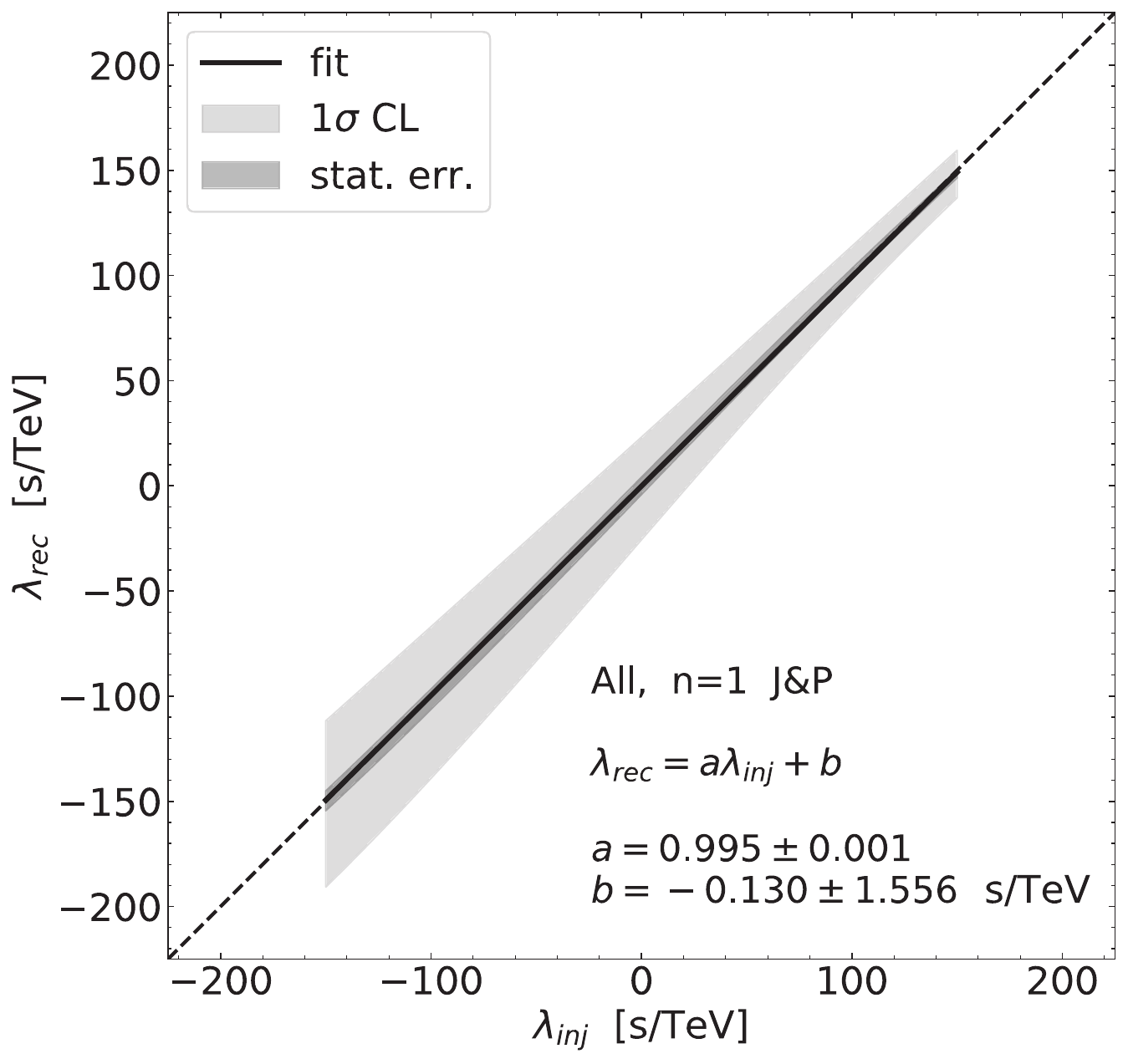}
    \caption{Calibration plot showing $\lambda_{rec}$ vs. $\lambda_{inj}$ for GRB~199114C (left) and all sources combined (right) in the linear case and J\&P formalism. The light gray area corresponds to the standard deviation of the $\lambda_{rec}$ distribution while the dark gray region shows the statistical uncertainty. For both plots, a function $a\ \lambda_{inj} + b$ is fitted (black line).}
    \label{fig:calib}
\end{figure*}

 The number of bins used in the tables has been chosen for each source to minimize the bias $\lambda_{rec} - \lambda_{inj}$ between the injected ($\lambda_{inj}$) and reconstructed ($\lambda_{rec}$) time delays. An example is shown in Figure~\ref{fig:convergence} for \object{GRB~190114C}. In this particular case, the plot shows that a minimum of $\sim$140~bins for each variable is required, and a conservative number of 200 was actually chosen. The optimal number of bins was found to be independent of the injected lag.
 Four sets of tables have been produced for each source which accounts for the four configurations explored in this work: J\&P or DSR formalism for distance, for linear and quadratic LIV effects.

In order to assess the sensitivity and precision of the lag reconstruction, simulated data sets were produced with different values for $\lambda_{inj}$. For each value of the injected lag, one thousand realizations of the light curve were simulated. The distribution of reconstructed values of $\lambda_n$ is shown in the central panel of Figure~\ref{fig:LL_UL_distribs} for \object{GRB~190114C}, in the J\&P case and $n = 1$. Lower and upper limits of the confidence intervals are taken as the values of $\lambda_n$ for which $2\,[L_S(\lambda_n)-\mathrm{min}(L_S)]=1$ for 68\% CL and $2\,[L_S(\lambda_n)-\mathrm{min}(L_S)]=3.84$ for 95\% CL (Equation~\ref{eq:LikelihoodData}). Their distributions for 68\% CL are displayed in the left and right panels of Figure~\ref{fig:LL_UL_distribs} respectively. All three distributions were fitted with asymmetric Gaussian functions providing three parameters: the average ($\lambda_{LL}$, $\lambda_{rec}$ and $\lambda_{UL}$) and standard deviations separately defined on the left and on the right of the maxima ($\sigma_{\lambda,l}$, $\sigma_{\lambda,r}$). While the latter accounts for statistical uncertainties only, the lower and upper limits $\lambda_{LL}$ and $\lambda_{UL}$ account for both statistical and systematic uncertainties.

\begin{figure*}[t]
    \plotone{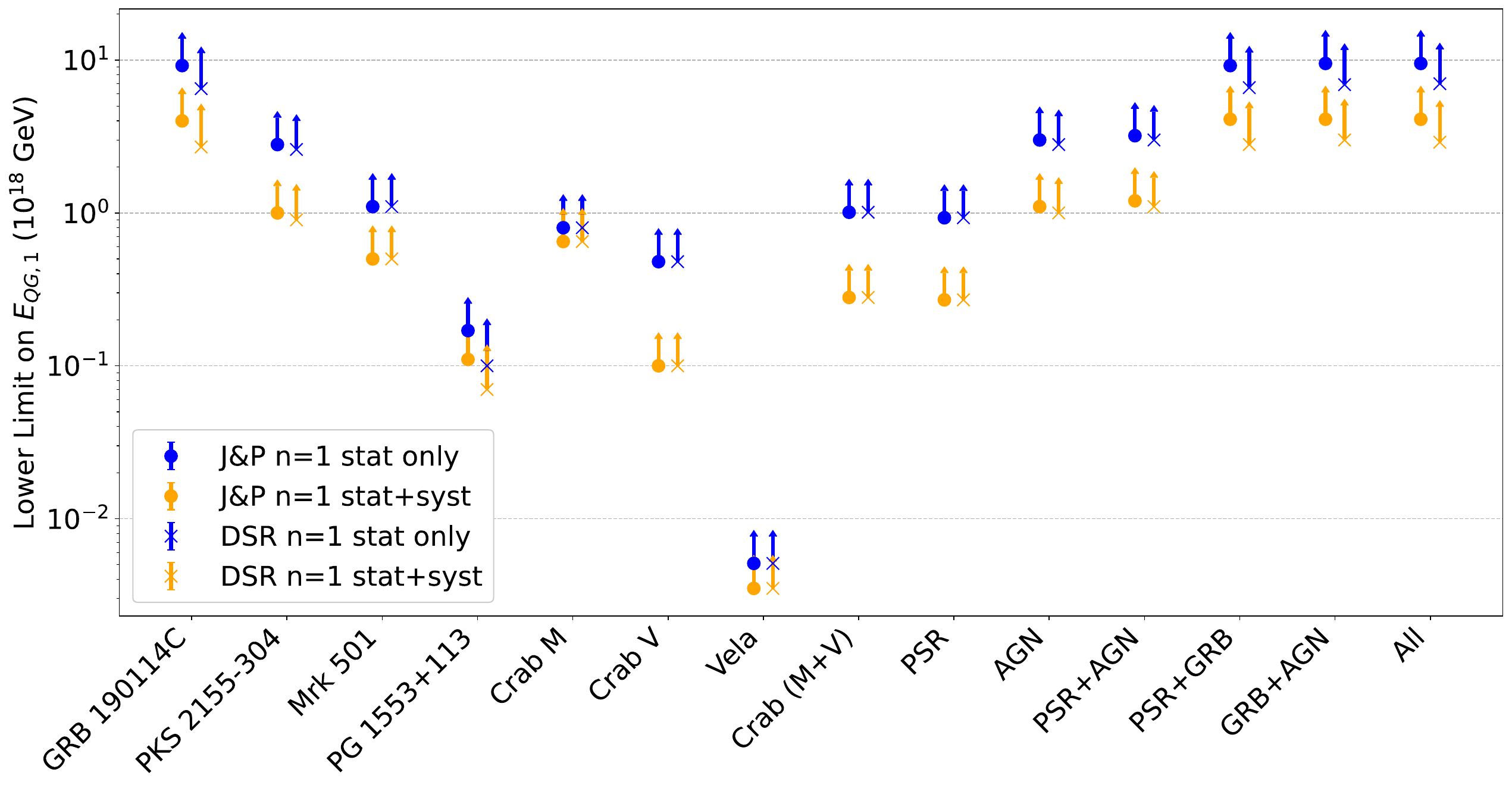}
    \plotone{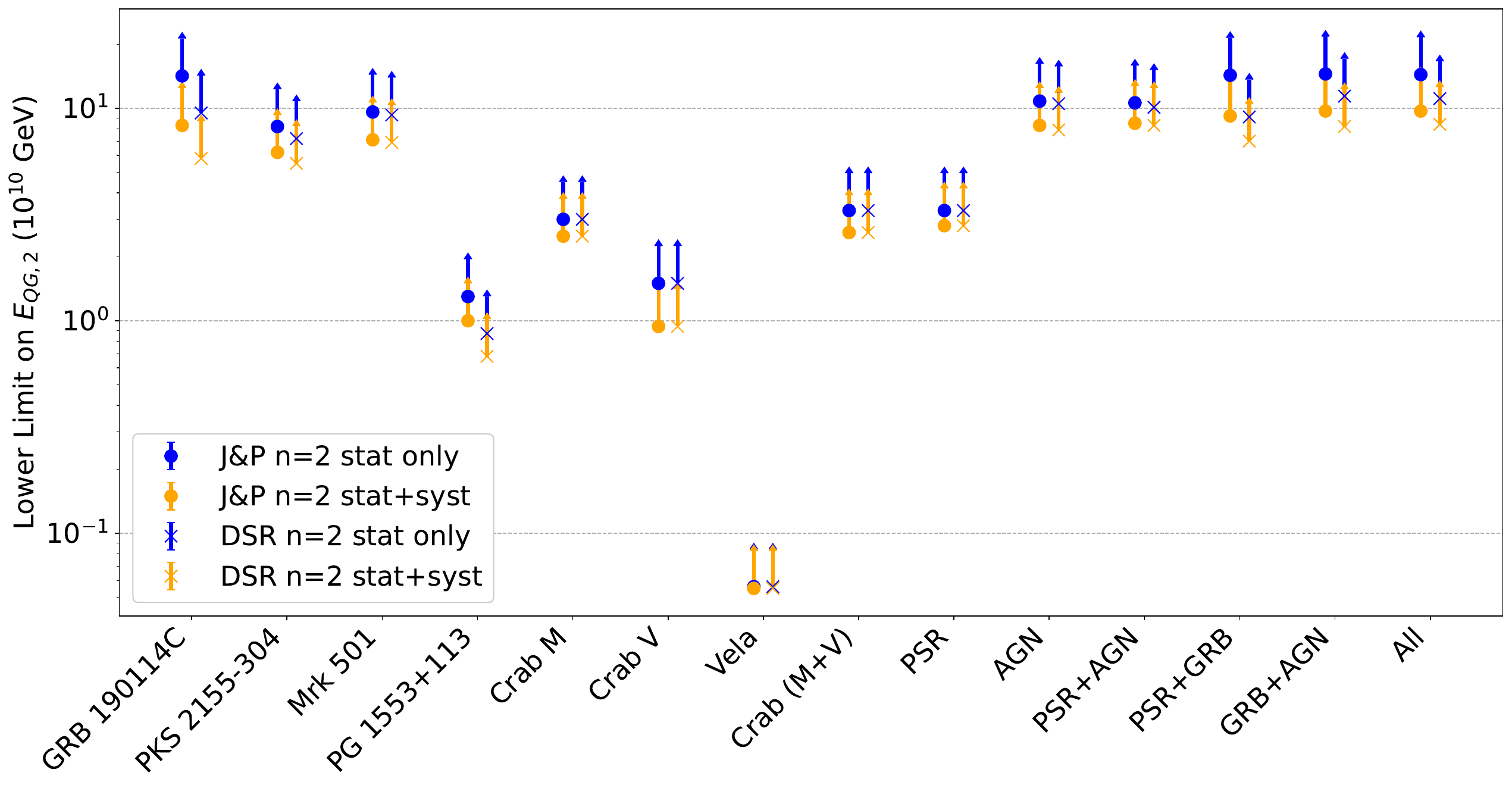}
    \caption{Limits obtained  for all individual sources and combinations for the linear (top) and quadratic (bottom) cases, for the J\&P (dots) and DSR (crosses) redshift dependence. Blue markers correspond to the case where only statistical errors are taken into account (``stat only'') while orange markers correspond to the case where both statistical and systematic errors are included (``stat+syst'').}
    \label{fig:result-comb}
\end{figure*}

For extragalactic sources, the range for $\lambda_{inj}$ goes from $-5\,\sigma_0$ to $+5\,\sigma_0$, where $\sigma_0 = \mathrm{max}(\sigma_{\lambda_{rec},l}, \sigma_{\lambda_{rec},r})$ for $\lambda_{inj} = 0$. In the case of PSR, the range for $\lambda_{inj}$ is chosen so that the highest energy photons are not shifted out of the phase range given in Table~\ref{tab:simulation}.

The plots for $\lambda_{rec}$ versus $\lambda_{inj}$ were then produced for individual sources as well as for combinations, for the two correction orders and the two lag-distance models. Figure~\ref{fig:calib} shows two examples of calibration plots for \object{GRB~190114C} alone (left) and for all sources combined (right) in the linear and J\&P case. The reconstructed lag is fitted with a linear function $\lambda_{rec} = a\ \lambda_{inj} + b$. The plot for the GRB alone shows a clear decrease of the reconstruction error as the injected lag increases. This is a consequence of a peculiar shape of the light curve, which has a narrow peak, followed by a power-law decay. As the value of $\lambda_{inj}$ increases, the peak of the light curve enters progressively the time window where the likelihood is computed, resulting in an improvement of the reconstruction precision.
The plot on the right shows the same behavior, illustrating how the GRB dominates over the other sources. Other examples of calibration plots are shown in Appendix~\ref{sec:annexB}. As none of them include the GRB, the consistency in the reconstruction error is maintained. All the plots produced show a very good reconstruction of the injected lag, with slopes $a$ very close to unity, while the bias $b$ is found to be close to zero.

\section{Results and discussion}\label{sec:res}

\subsection{Systematic uncertainties}

All systematic uncertainties are listed for individual sources and combinations in Table~\ref{tab:systematics_res} (J\&P case) and Table~\ref{tab:systematics_res_DSR} (DSR case) in the Appendix. For most of the individual sources, the dominant systematic is the statistical uncertainty of the light curve template. Since the time lag intensifies as the correction order $n$ gets larger, the template uncertainties contribute comparatively less in the quadratic case than the linear one. For other individual sources, the precision of the energy distribution of the events prevails. Indeed, the energy scale uncertainty is found to be the most important source of systematics for the Crab pulsar observed by MAGIC, the Vela pulsar and \object{Mrk~501}, for the quadratic case. This is expected since the time delay depends on the energy squared. A similar behavior is observed for \object{GRB~190114C} and the Crab pulsar observed by VERITAS, where the uncertainty on the spectral slope dominates.

\begin{table*}[t]
\caption{95\% CL limits obtained for individual objects and combinations.\label{tab:combined_res}}
\begin{center}
\scriptsize
\begin{tabular}{lrlrlrlrl}
\hline
\hline
Source & \multicolumn{4}{c}{$E_{QG,1}$} & \multicolumn{4}{c}{$E_{QG,2}$} \\
    & \multicolumn{2}{c}{J\&P}  & \multicolumn{2}{c}{DSR} & \multicolumn{2}{c}{J\&P} & \multicolumn{2}{c}{DSR} \\
    & \multicolumn{2}{c}{($10^{18}$ GeV)} & \multicolumn{2}{c}{($10^{18}$ GeV)} & \multicolumn{2}{c}{($10^{10}$ GeV)} & \multicolumn{2}{c}{($10^{10}$ GeV)} \\
    & w/o syst. & w/ syst. & w/o syst. & w/ syst. & w/o syst. & w/ syst. & w/o syst. & w/ syst.  \\
\hline
GRB 190114C & 9.2 & 4.0 & 6.5 & 2.7 & 14.2 & 8.3 & 9.5 & 5.8 \\
PKS  2155-304 & 2.8 & 1.0 & 2.6 & 0.9 & 8.2 & 6.2 & 7.2 & 5.5 \\
Mrk~501 & 1.1 & 0.5 & 1.1 & 0.5 & 9.6 & 7.1 & 9.3 & 6.9 \\
PG 1553+113 & 0.17 & 0.11 & 0.10 & 0.07 & 1.3 & 1.0 & 0.87 & 0.68 \\
Crab (M)   & 0.80 & 0.65 & - & - & 3.0 & 2.5 & - & - \\
Crab (V)   & 0.48 & 0.10 & - & - & 1.5 & 0.94 & - & - \\
Vela  & $5.1 \times 10^{-3}$ & $3.5 \times 10^{-3}$ & - & - & $5.6 \times 10^{-2}$ & $5.5 \times 10^{-2}$ & - & - \\
\hline
Crab (M+V)  & 1.0 & 0.28 & - & - & 3.3 & 2.6 & - & - \\
PSR & 1.0 & 0.28 & - & - & 3.3 & 2.8 & - & - \\
AGN & 3.0 & 1.1 & 2.8 & 1.0 & 10.8 & 8.3 & 10.5 & 7.9 \\
AGN+PSR & 3.2 & 1.2 & 3.0 & 1.1 & 10.6 & 8.5 & 10.1 & 8.3 \\
GRB+PSR & 9.2 & 4.1 & 6.6 & 2.8 & 14.3 & 9.2 & 9.1 & 7.0 \\
GRB+AGN & 9.5 & 4.1 & 6.9 & 3.0 & 14.5 & 9.7 & 11.4 & 8.2 \\
\hline
All combined & 9.5 & 4.1 & 7.0 & 2.9 & 14.4 & 9.7 & 11.1 & 8.4 \\
\hline%
\end{tabular}%
\end{center}%
\end{table*}

For the combinations, dominant systematic uncertainties are the ones of sources that dominate the sample. The pulsar combinations are dominated by template statistics, while the combination of AGN shows a predominance of template statistics for $n=1$, and a domination of energy scale for $n=2$, confirming the importance of the energy distribution uncertainty for the quadratic case. The combinations that include \object{GRB~190114C} follow a very similar trend due to the dominance of the GRB over the other sources. They show a clear ascendancy of the uncertainty of the power law index, which is the main source of systematic uncertainty for the combination of all the sources.

\begin{figure*}[t]
    \plottwo{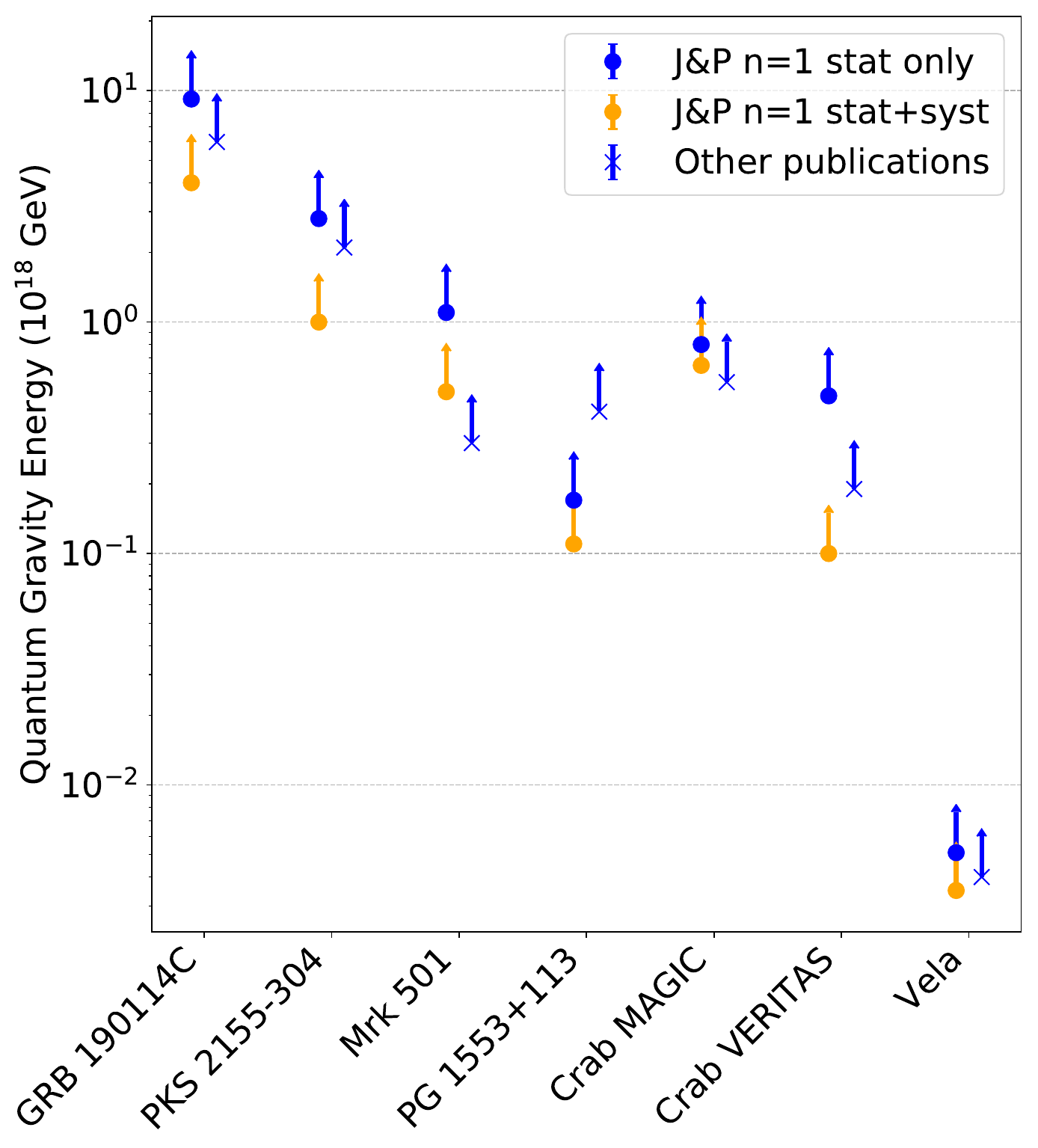}{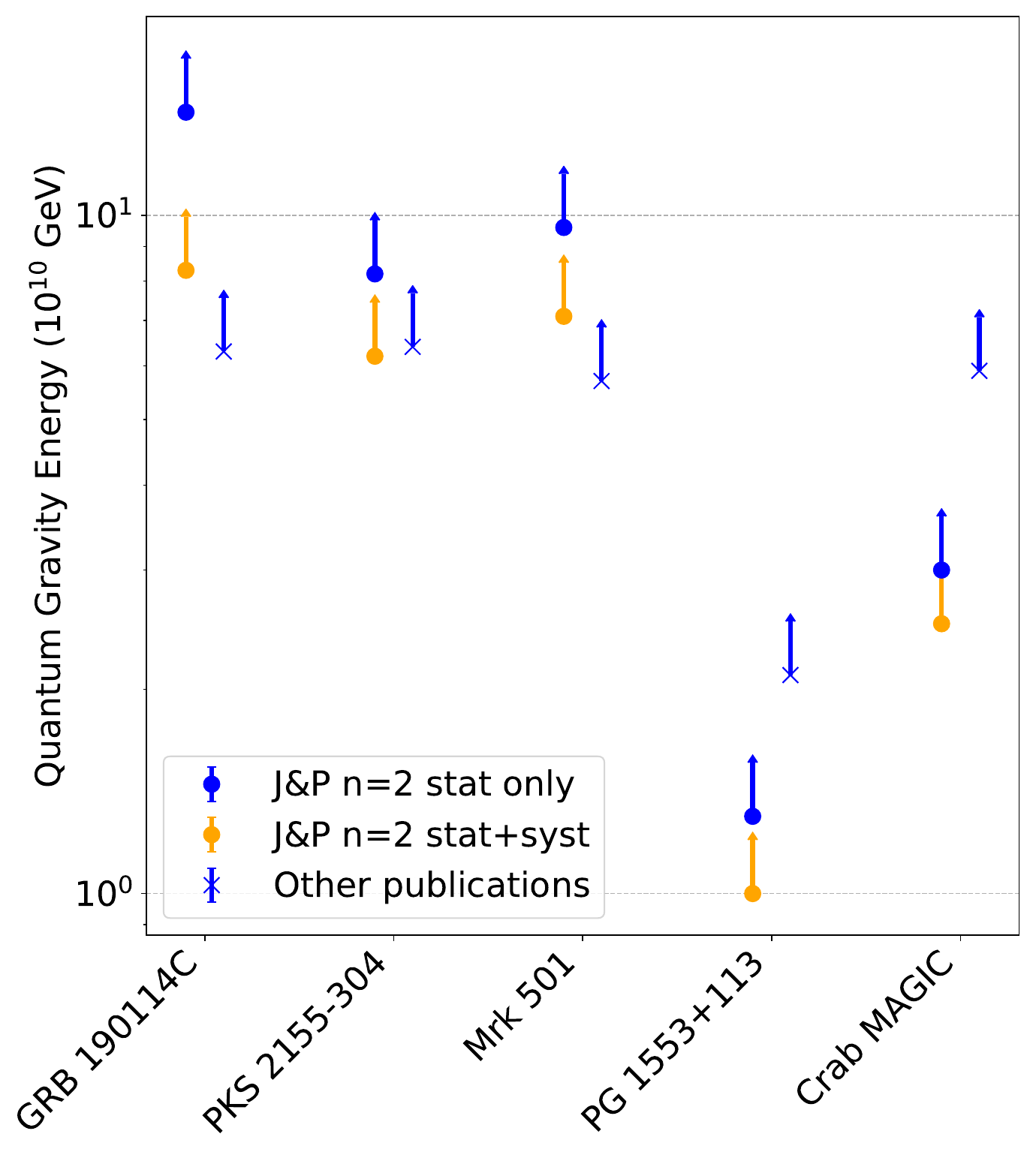}
    \caption{Limits obtained from the simulated data sets for all individual sources for the linear (left) and quadratic (right) cases for the J\&P redshift dependence. Blue dots show the limits obtained taking into account statistical errors only (``stat only'') while yellow dots show the limits including both statistical and systematic errors (``stat+syst''). Blue crosses give the limits published from actual data sets (Table~\ref{tab:all_res}).}
    \label{fig:result-1}
\end{figure*}

\subsection{Limits}

\begin{figure*}[t]
    \plottwo{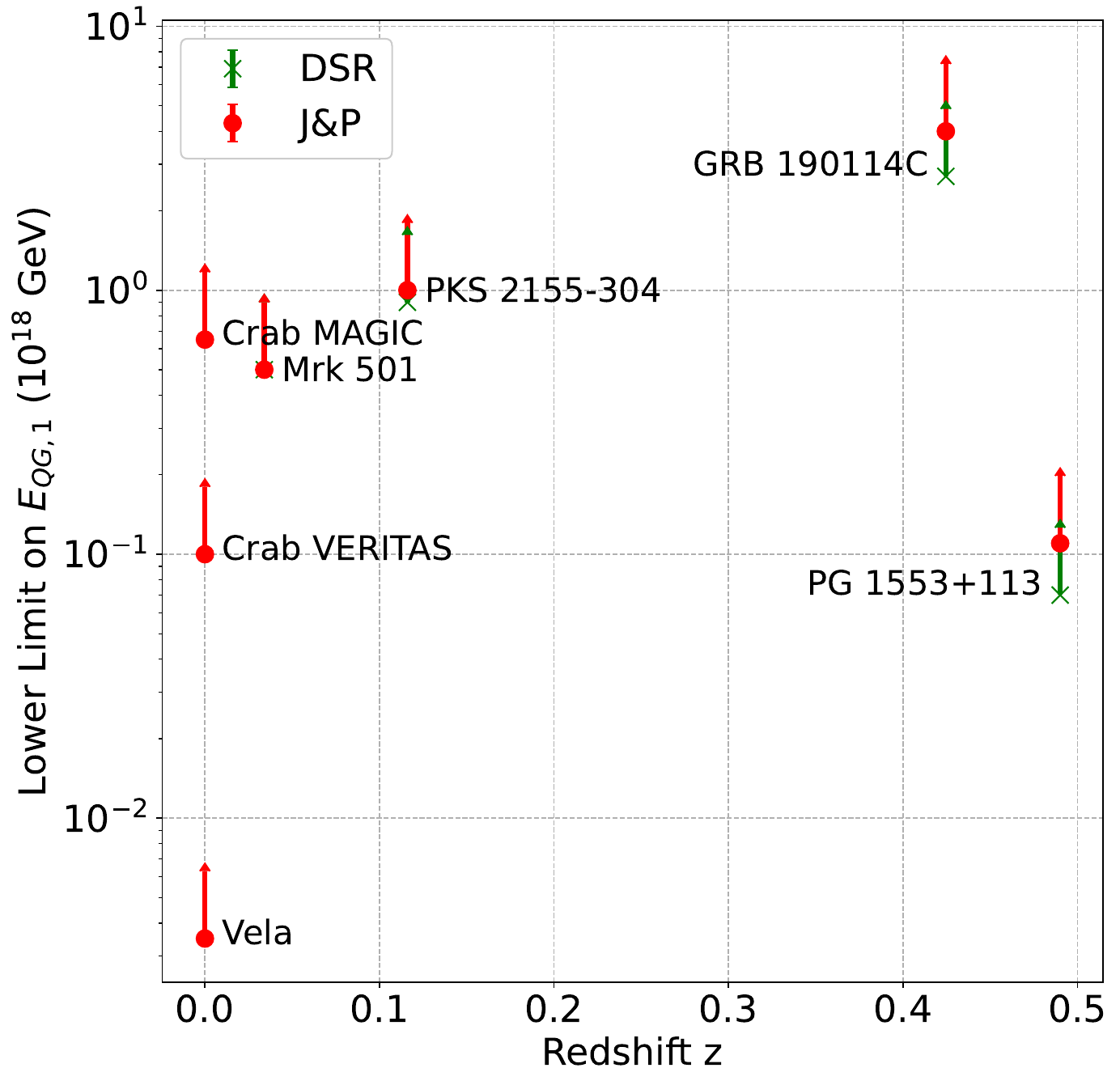}{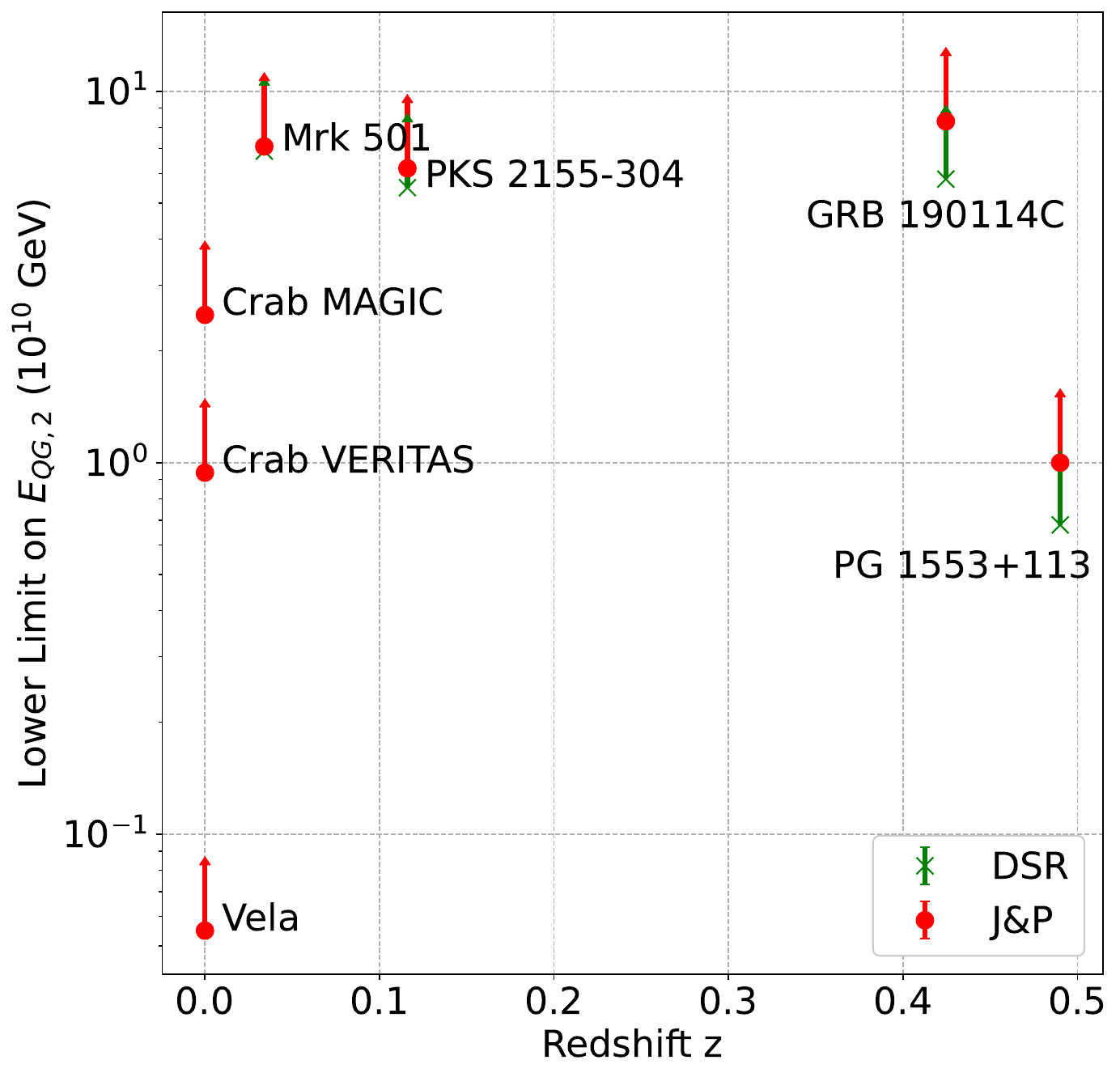}
    \caption{Comparison between the limits obtained in the J\&P framework (red dots) and the limits obtained in the DSR formalism (green crosses) in the linear case (left) and quadratic case (right). The limits shown include both statistical and systematic errors.}
    \label{fig:result-3}
\end{figure*}

From equation~(\ref{eq:lambda}), limits on $E_{QG}$ are given by:
\begin{equation}
\label{eq:limits}
\left[{ \frac{2}{n+1} \left( \lambda_{n,\pm} + \sqrt{\delta_{stat}^2 + \sigma^2\delta_{syst}^2} \right) \mathrm{H}_\mathrm{0}} \right] ^{\frac{1}{n}},
\end{equation}
where the subscript $\pm$ refers to subluminal and superluminal cases, $\delta_{stat}$ is the statistical error (standard deviation) on the normally distributed reconstructed value of $\lambda_{n}$, $\delta_{syst}$ is the overall systematic error obtained from the values listed in Tables \ref{tab:systematics_res} and \ref{tab:systematics_res_DSR} computed for a confidence level of 68\%, and $\sigma$ is a real number allowing for a shift in confidence level using the same systematic errors. Since systematic errors are computed for 68\%~CL, and statistical errors for 95\% CL, $\sigma$ is set to two in the following.

Using Equation~(\ref{eq:limits}), $E_{QG}$ limits were obtained for both subluminal and superluminal cases. Both approaches give comparable results and only the subluminal limits are shown in Table \ref{tab:combined_res}. In Figure \ref{fig:result-comb}, results are given with and without accounting for systematic uncertainties, clearly demonstrating the importance of taking them into account. In some cases, systematics lead to upper limits smaller by a factor of $\gtrsim 2$ as compared to the case they are not taken into account. For pulsars, DSR and J\&P formalisms lead to the same limits so they are given only for the J\&P case in the table.

Figure~\ref{fig:result-1} shows a comparison between the already published results taken from the references listed in Table~\ref{tab:all_res} and the ones obtained in the present study, for $n=1$ (left) and $n=2$ (right). Overall, the agreement between simulations and data is good, showing the simulated data sets represent the actual data well. The observed differences are most probably due to three different factors. First, the limits obtained in this work come from several hundreds realizations of the light curves while already published limits were derived from one (measured) light curve. Second, the systematic uncertainties in previous publications were evaluated with different methods. These methods can vary from one analysis to another, but most use a frequentist approach, while in the present paper nuisance parameters and profile likelihood where used (Section~\ref{subsec:uncertainties}). Finally, IRFs were fully taken into account in the present analysis while it was often approximated as a constant of energy in earlier articles. The latter point was fully justified at the time by the use of a somewhat reduced energy range, while we wanted to get rid of this restriction in the present analysis.

As expected from previously published results, GRB~190114C is the most constraining source due to its high redshift, high variability, large statistics, as well as the fact it has been observed on a wide energy range. Therefore, it dominates the final result whenever it is included in the combination.

When the GRB observation is not included, AGNs dominate with a competition between \object{PKS~2155-304} and \object{Mrk~501}. Due to its smaller number of events and limited energy range, \object{PG~1553+113} limit is less constraining, even though its redshift is the highest of all the sources included in this work. While \object{PKS~2155-304} dominates over the other sources in the linear case due to its higher redshift and event statistics, \object{Mrk~501} dominates the limit in the quadratic case due to its energy range extending twice as high as the one of \object{PKS~2155-304}.

PSR have only a marginal impact on the overall combination due to their closeness. The Crab pulsar dominates the combined PSR limit thanks to its higher statistics, wider energy range and greater distance. However, it is important to note that the limits provided by pulsars are independent of the redshift dependence model, providing model free constraints.

Figure~\ref{fig:result-3} shows the limits on $E_{QG,n}$ as a function of the redshift for both DSR and J\&P models. Differences in the results from the two approaches start to become significant for high redshift sources such as \object{GRB~190114C} or \object{PG~1553+113}, hence behaving in accordance with the $\kappa_n$ parameter evolution shown in Figure~\ref{fig:kappaz}. Due to the facts that $\kappa^\mathrm{J\&P} > \kappa^\mathrm{DSR}$ and $\kappa^\mathrm{J\&P}$ increases faster than $\kappa^\mathrm{DSR}$ (Section~\ref{sec:dist}), the J\&P model emphasizes the impact of large redshift sources on the limits. Therefore, the GRB dominates more in the J\&P case than for the DSR case, where all sources contributions are more balanced.

\section{Conclusions}

In the present paper, we have described an implementation of likelihood analysis designed with the goal to combine data from different sources and experiments in the search for LIV-induced energy-dependent time delays. One of the most important benefits of the likelihood technique is its simplicity for such a combination. In order to check the method and evaluate its performance, simulated data sets were produced mimicking actual observations of one GRB, three flaring AGN and two pulsars by the H.E.S.S., MAGIC and VERITAS experiments. We paid particular attention to the implementation of the algorithm, checking for any bias and carefully evaluating statistical and systematic errors, and their combination within the different experiments. For the first time, two different formalisms were studied concerning the way the distance is taken into account in the time-lag computation. Others could be added in the future \citep[see e.g.][]{Amelino2021}. As the next step, the software developed for this work will be applied to all available data sets recorded so far by H.E.S.S., MAGIC and VERITAS, and perhaps by other experiments, and the results will be published in the second part of this work.

Another important advantage of likelihood analysis is its adaptability. Indeed, nothing prevents, in principle, including other effects on production or propagation of photons in the probability density function. Two examples can be pointed out. First, as mentioned in the introduction, it is known that LIV could modify the absorption of VHE photons by the EBL changing the shape of high energy spectra. Assuming that QG affects both the photon group velocity and photon interactions, the likelihood technique could be used to provide combined EBL and delay constraints on QG models. It is not clear however, whether the different effects would manifest at the same energy scale, or if a different energy scale is applicable for each effect.
Second, it should be possible to include other types of delays in the probability function to probe both propagation and source-intrinsic time lags. Despite some recent exploratory work on that topic \citep[see e.g.][for the case of blazar flares]{Perennes2020}, the latter are still poorly understood. In addition, intrinsic effects are most probably different from one type of source to another and even from one sub-type to another: short or long GRBs, blazars or flat spectrum radio quasars. We therefore chose not to include them in the present study. Intrinsic effects are a critical aspect and they will need to be addressed in the future.

The Cherenkov Telescope Array\,\footnote{\url{https://www.cta-observatory.org}} (CTA) will start operating soon, superseding the current generation IACTs in the years 2025-2030 \citep{Acharya2019}. Thanks to its better overall performance and dedicated observation strategies to maximize the number of transient event detection, it is expected that both CTA arrays (one in each hemisphere) will be able to detect a large number of PSR, AGN flares and GRB. Different sub-array configurations will be used in order to optimize the observation program and combining data will therefore become very important. {As a result, CTA will be much more sensitive to LIV effects than current generation experiments.} The tools developed in this work will be made publicly available concurrently with the publication of the second paper and adapted to be used in CTA analysis software architecture.

\begin{acknowledgments}

The authors would like to thank collabora\-tions H.E.S.S., MAGIC and VERITAS for their support in the making of this joint effort and for allowing the use of IRFs for the set of sources used in this paper. They also would like to acknowledge networking support by the COST Action CA18108 (\url{https://qg-mm.unizar.es/}). This project has received funding from the European Union's Horizon 2020 research and innovation programme under the Marie Sk\l{}odowska-Curie grant agreement No. 754510, from the ERDF under the Spanish Ministerio de Ciencia e Innovaci\'{o}n (MICINN), grant PID2019-107847RB-C41, the Centro de Excelencia ``Severo Ochoa'' (SEV-2016-0588) and from the CERCA program of the Generalitat de Catalunya.
T.T. acknowledges funding from the University of Rijeka, project number uniri-prirod-18-48, and from the Croatian Science Foundation (HrZZ), project number IP-2016-06-9782.

The authors would like to thank G. D'Amico for his useful comments on the draft as well as G. Rosati and C. Pfeifer for insightful discussions on lag-redshift dependence. {Finally, the authors express their gratitude to the anonymous referee who helped clarifying some parts of the paper.}

This paper is dedicated to the memory of our colleague and friend A. Jacholkowska, who initiated this work and put it on the best tracks towards a successful completion.

\end{acknowledgments}

\facilities{HESS,MAGIC,VERITAS}
\software{ROOT}

\newpage
\appendix
\restartappendixnumbering

\section{Systematic uncertainties}

All systematic uncertainties are listed for individual sources and combinations in Table~\ref{tab:systematics_res} (J\&P case) and Table~\ref{tab:systematics_res_DSR} (DSR case).

\begin{table*}[h!]
\begin{center}
\caption{Summary of systematic uncertainties for all sources and combinations simulated for the J\&P case.\label{tab:systematics_res}}
\scriptsize
\begin{tabular}{lccccccccc}
\hline
\hline
Source & Correction & Template & Energy & Background & Uncertainty on  & Distance/redshift & Reconstruction & All syst. \\
       &  order      & statistics & scale & normalization & power law index & uncertainty & uncertainty & combined \\
       &        & (s.TeV$^{-n}$) & (s.TeV$^{-n}$) & (s.TeV$^{-n}$) & (s.TeV$^{-n}$) & (s.TeV$^{-n}$) & (s.TeV$^{-n}$) & (s.TeV$^{-n}$) \\
\hline
\multirow{2}{*}{GRB 190114C} & $n = 1$ & $17.8$ & $6.9$ & $8.0$   & $9.4$   & $< 7.7$ & $3.0$ & $25.6$   \\ 
            &  $n = 2$ & $9.4$   & $12.4$   & $1.7$   & $15.4$   & $<9$ & $4.2$ & $24.1$   \\ 
\multirow{2}{*}{PKS  2155-304} & $n = 1$ & $101$   & $11.7$   & $<20$   & $<22$   & $17.8$ & $< 3.3$  & $107$   \\ 
            &  $n = 2$ & $21.8$ & $19.3$ & $0.7$& $8.1$& $12.0$ & $< 2.2$  & $37.4$  \\ 
\multirow{2}{*}{Mrk~501} & $n = 1$ & $155$ & $56$ & $< 51$ & $49$ & $1.$ &  $< 8.5$ & $197$\\ 
            &  $n = 2$ & $11.2$ & $18.3$ & $<10.3$ & $9.3$ & $0.19$ & $ < 1.6$  & $28.8$\\ 
\multirow{2}{*}{PG1553+113} & $n = 1$ & $631$    & $150$    & $324$    & $<361$    & $112$  & $< 64$ & $727$   \\ 
            &  $n = 2$ & $916$   & $638$   & $537$   & $< 552$   & $338$   & $< 112$ & $1282$  \\ 
\multirow{2}{*}{Crab V} & $n = 1$ & $897$ & $137$ & $<73$ & $142$ & $145$ & $<25$ & $1135$ \\ 
            &  $n = 2$ & $1141$ & $410$ & $<264$ & $694$ & $265$ & $<174$ & $1820$\\ 
\multirow{2}{*}{Crab M} & $n = 1$ & $371$ & $66$ & $ 7$ & $23$ & $74$ & $<11$ & $416$\\ 
            &  $n = 2$ & $167$ & $64.5$ & $61$ & $24$ & $48$ & $<72$ & $190$\\ 
\multirow{2}{*}{Vela} & $n = 1$ & $1.36 \times 10^4$   & $ 1.03 \times 10^4$   & $0.46 \times 10^4$   &  $< 1.3 \times 10^4$   &  $ 1.30 \times 10^3$ & $< 5.87 \times 10^3$  & $ 2.28 \times 10^4$  \\ 
            &  $n = 2$ & $1.0 \times 10^5$   & $2.05 \times 10^5$   & $0.48 \times 10^5$   &  $ < 1.5 \times 10^5$   &  $ 1.57 \times 10^5$ &  $<0.95 \times 10^5$ & $ 3.05 \times 10^5$  \\ 
\hline
\multirow{2}{*}{Crab (M+V)} & $n = 1$ & $357$ & $49$ & $<56$ & $32$ & $61$ & $<32$ & $398$\\ 
            &  $n = 2$ & $161$ & $59$ & $45$ & $59$ & $38$ & $<83$ & $197$\\ 
\multirow{2}{*}{PSR} & $n = 1$ & $355$ & $52$ & $< 58$ & $38$ & $58$ & $<11$ & $394$ \\ 
            &  $n = 2$ & $90$ & $71$ & $49$ & $24$ & $62$ & $<55$ & $138$ \\ 
\multirow{2}{*}{AGN} & $n = 1$ & $89.5$ & $12$ & $ <15 $& $3.7$& $15.8$ & $<2.9$  & $94.9$\\
            &  $n = 2$ & $10.1$ & $11.1$ & $<6$ & $6.2$ & $3.4$ & $<1.3$ & $19.7$\\
\multirow{2}{*}{AGN+PSR} & $n = 1$ & $85$ & $11$ & $< 18$ & $5$ & $15$ & $<2.9$ & $91$ \\ 
            &  $n = 2$ & $9.6$ & $10.9$ & $<8$ & $5.9$ & $4.5$ & $<1.1$ & $17.8$ \\ 
\multirow{2}{*}{GRB+AGN} & $n = 1$ & $17.8$ & $5.8$ & $6.8$& $8.3$ & $1.4$ & $3.3$ & $24.5$\\
            &  $n = 2$ & $6.8$ & $7.8$ & $<6.6$ & $9.0$ & $1.7$ & $1.4$ & $16.2$\\ 
\multirow{2}{*}{GRB+PSR} & $n = 1$ & $17.5$ & $6.7$ & $7.9$ & $9.1$ & $1.0$ & $3.2$ & $24.9$ \\ 
            &  $n = 2$ & $8.1$ & $11.3$ & $1.6$ & $12.7$ & $2.8$ & $< 1.1$ & $19.4$ \\ 
\multirow{2}{*}{All} & $n = 1$ & $18.0$ & $5.8$ & $6.7$ & $8.2$ & $1.5$ & $4.1$ & $24.8$ \\ 
            &  $n = 2$ & $7.5$ & $7.7$ & $< 6.2$ & $8.2$ & $2.4$ & $4.8$ & $16.4$ \\ 
\hline
\end{tabular}
\end{center} 
\end{table*}

\begin{table*}[h!]
\begin{center}
\caption{Summary of systematic uncertainties for all sources and combinations simulated for DSR case.\label{tab:systematics_res_DSR}}
\scriptsize
\begin{tabular}{lccccccccc}
\hline
\hline
Source & Correction & Template & Energy & Background & Uncertainty on  & Distance/redshift & Reconstruction & All syst. \\
       &  order      & statistics & scale & normalization & power law index & uncertainty & uncertainty & combined \\
       &        & (s.TeV$^{-n}$) & (s.TeV$^{-n}$) & (s.TeV$^{-n}$) & (s.TeV$^{-n}$) & (s.TeV$^{-n}$) & (s.TeV$^{-n}$) & (s.TeV$^{-n}$) \\
\hline
\multirow{2}{*}{GRB 190114C} & $n = 1$ & $26.2$  & $10.2$  & $11.9$  & $13.9$  & $<11.2$  & $5.5$ & $38.0$  \\ 
            &  $n = 2$ & $18.0$  & $25.5$  & $3.8$  & $30.0$  & $6.2$  &  $10.6$ & $47.8$  \\ 
\multirow{2}{*}{PKS  2155-304} & $n = 1$ & $113$  & $12.7$  & $< 22.5 $  & $< 24.2 $   & $17.3$  & $<3.6$  & $119$ \\ 
            &  $n = 2$ & $25.8$  & $23.7$  & $3.4$  & $7.0$  & $14.8$  & $<2.9$ & $45.6$ \\
\multirow{2}{*}{Mrk~501} & $n = 1$ & $160$ & $58$ & $< 53$ & $51$ & $1.$& $<8.0$  & $204$\\ 
            &  $n = 2$ & $12.0$ & $19.6$ & $<11$ & $10.0$ & $0.2$ & $<1.8$  & $30.9$\\
\multirow{2}{*}{PG1553+113} & $n = 1$ & $968$ & $311$ & $545$ & $<555$ & $<522$ & $<104$ & $1131$\\
            & $n = 2$ & $2200$ & $1545$ & $1259$ & $<1377$ & $295$ & $<250$ & $2965$\\
\hline
\multirow{2}{*}{AGN} & $n = 1$ & $98.4$ & $12.9$ & $<17$ & $4.2$& $14.8$& $<3.2$ & $103$\\
            &  $n = 2$ & $11.1$ & $13.0$ & $<6.6$ & $7.3$ & $2.1$ & $<1.5$  & $22.5$\\
\multirow{2}{*}{AGN+PSR} & $n = 1$ & $94$ & $12$ & $<19$ & $4.3$ & $15$ & $<3.0$ & $99$\\ 
            &  $n = 2$ & $9.1$ & $11.9$ & $<8.2$ & $6.1$ & $3.9$ & $<1.2$ & $19.1$ \\ 
\multirow{2}{*}{GRB+AGN} & $n = 1$ & $26.2$ & $7.7$ & $9.1$ & $11.2$ & $2.4$ & $1.7$ & $34.7$ \\ 
            &  $n = 2$ & $10.1$ & $11.2$ & $<8.5$ & $9.8$ & $1.7$& $4.3$  & $21.7$\\
\multirow{2}{*}{GRB+PSR} & $n = 1$ & $26.0$ & $9.7$ & $11.3$ & $13.3$ & $1.8$ & $3.9$ & $37.4$ \\ 
            &  $n = 2$ & $8.0$ & $18.0$ & $<15.4$ & $18.5$ & $6.5$ & $<2.5$ & $28.7$ \\ 
\multirow{2}{*}{All} & $n = 1$ & $27.0$ & $7.7$ & $8.7$ & $10.9$ & $2.8$ & $<4.5$ & $35.6$ \\ 
            &  $n = 2$ & $10.1$ & $11.0$ & $<0.96$ & $8.3$ & $3.2$ & $<4.2$ & $19.8$ \\ 
\hline
\end{tabular}
\end{center} 
\end{table*}

\section{Calibration plots for combined AGN and combined PSR}\label{sec:annexB}

\begin{figure*}[h]
    \centering
    \plottwo{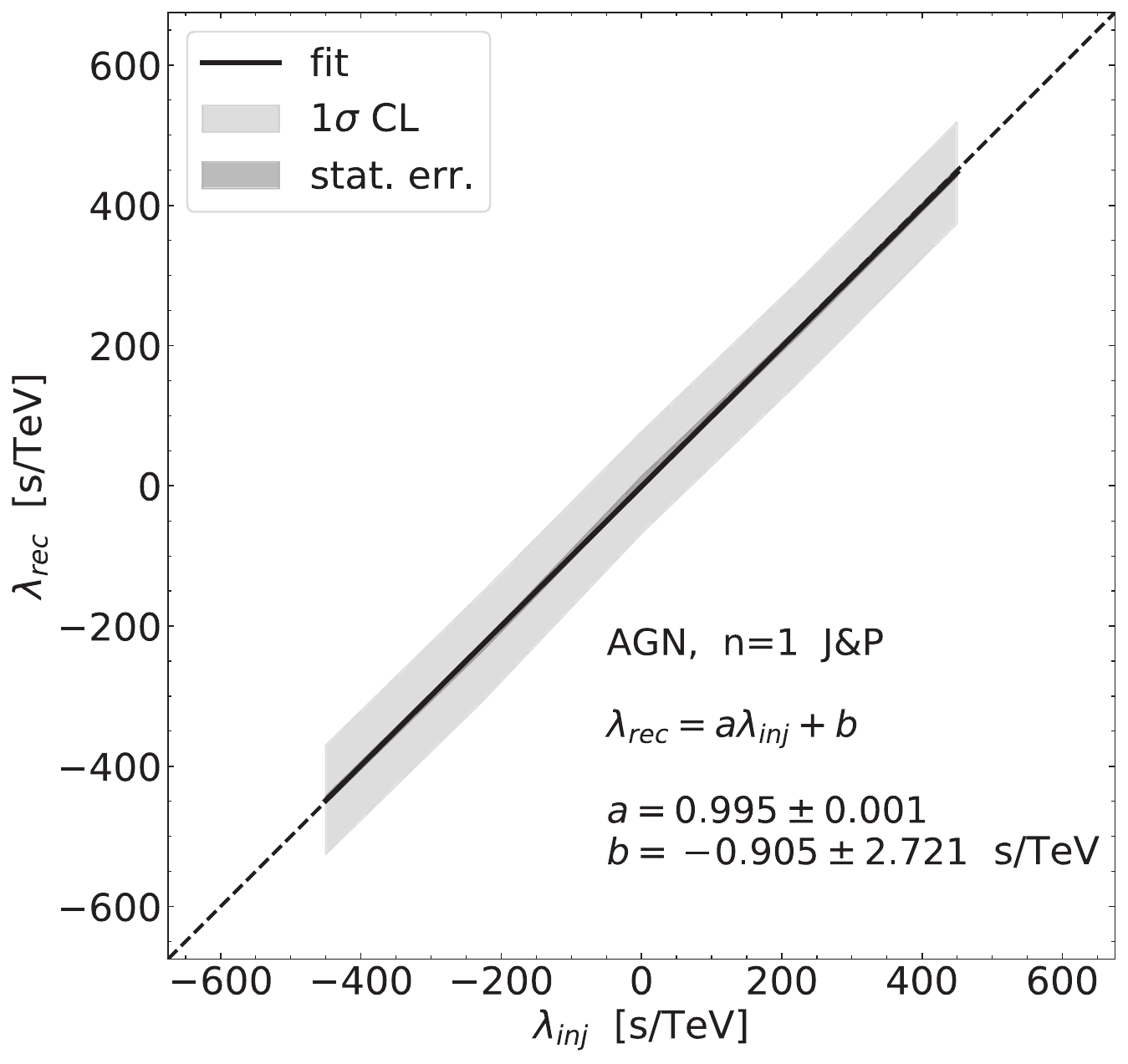}{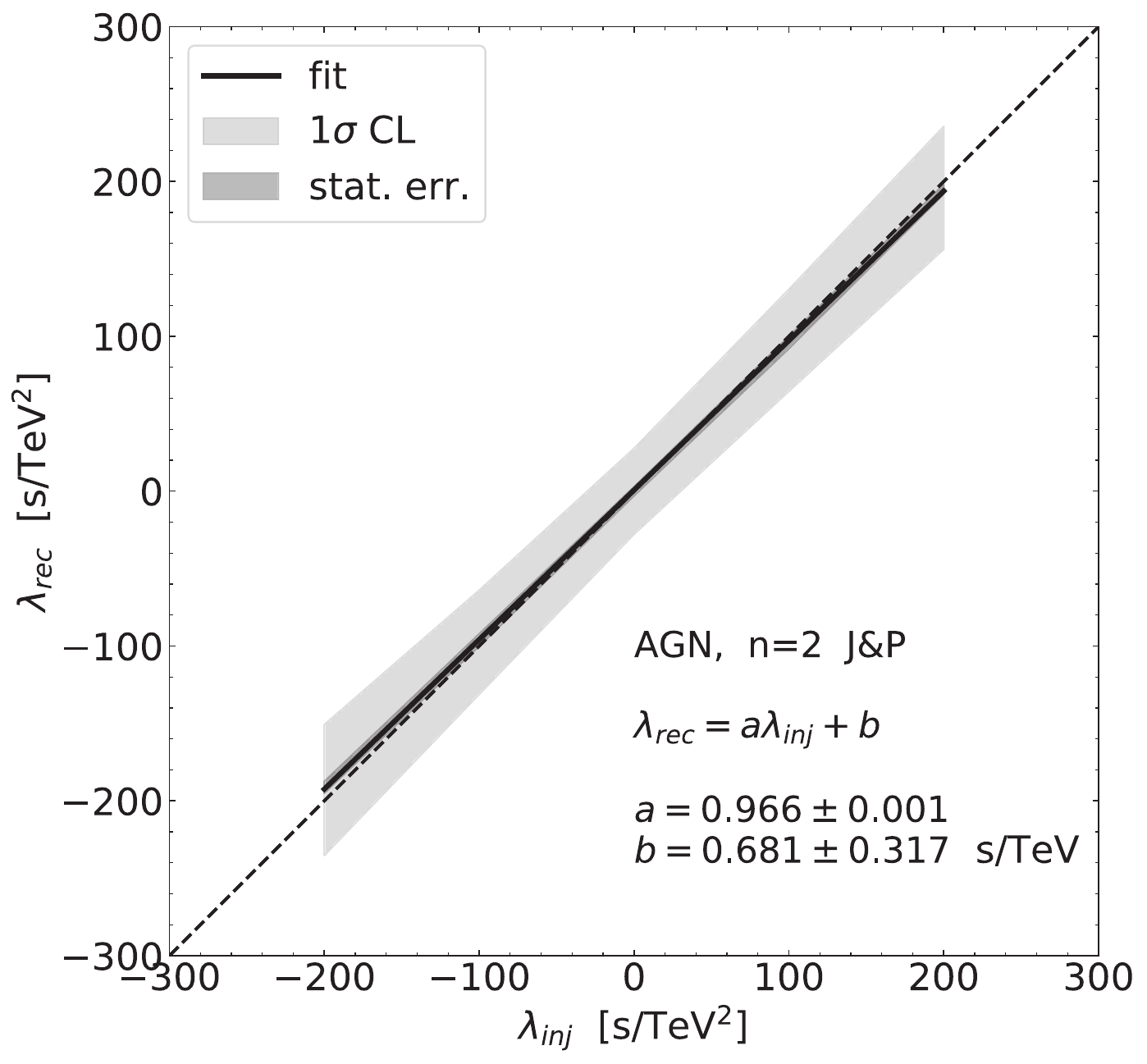}
    \caption{Calibration plots showing $\lambda_{rec}$ vs. $\lambda_{inj}$ for all AGN combined for the linear case (left) and the quadratic case (right) with the J\&P formalism. The light gray area corresponds to the standard deviation of the $\lambda_{rec}$ distribution while the dark gray region shows the statistical uncertainty. For both plots, a function $a\ \lambda_{inj} + b$ is fitted (black line).}
    \label{fig:comb_AGN_n1}
\end{figure*}

\begin{figure*}[h]
    \centering
    \plottwo{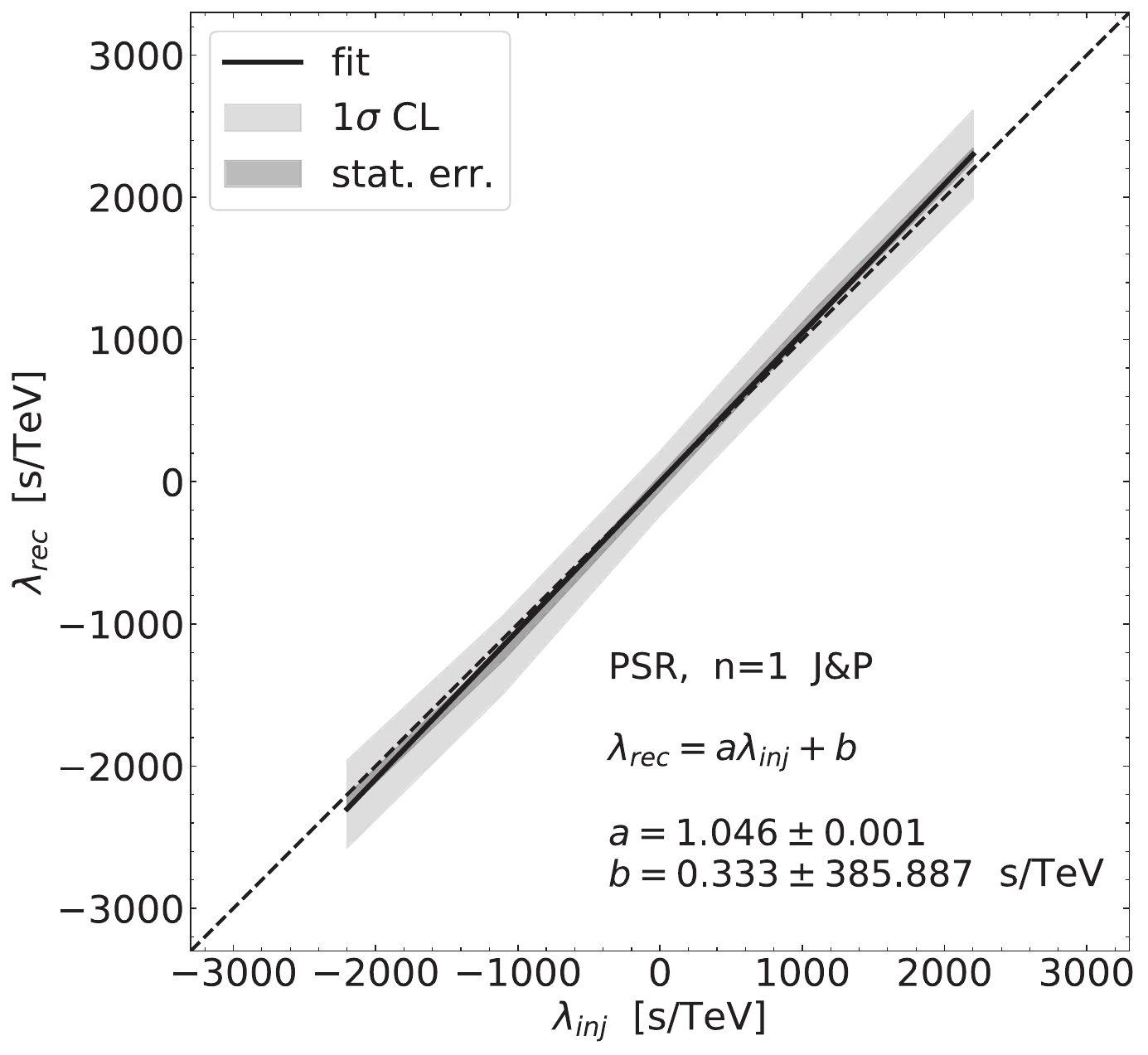}{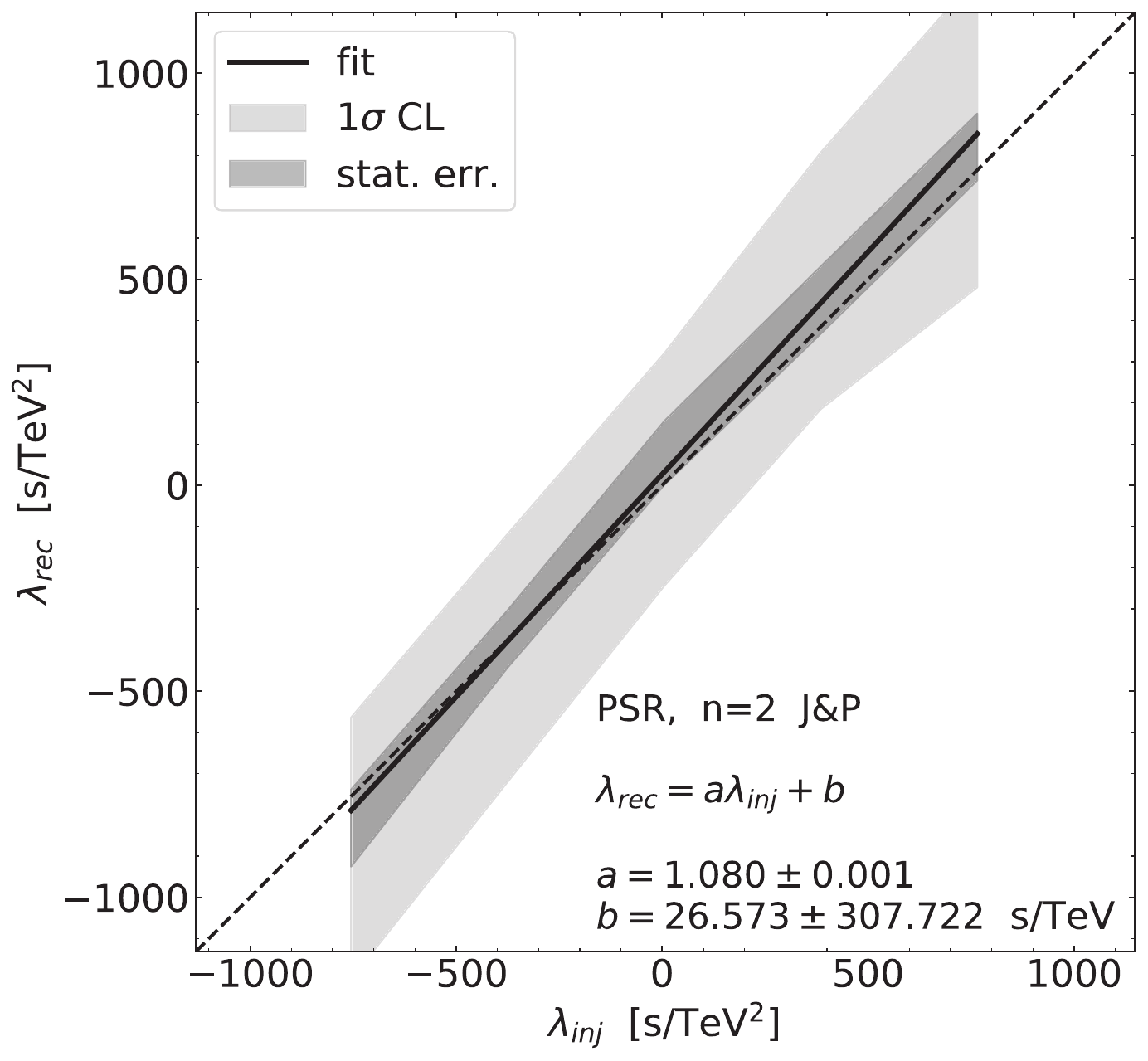}
    \caption{Calibration plots showing $\lambda_{rec}$ vs. $\lambda_{inj}$ for all PSR combined for the linear case (left) and the quadratic case (right) with the J\&P formalism. The light gray area corresponds to the standard deviation of the $\lambda_{rec}$ distribution while the dark gray region shows the statistical uncertainty. For both plots, a function $a\ \lambda_{inj} + b$ is fitted (black line). Note that the scales are not the same for $n=1$ and $n=2$.}
    \label{fig:comb_PSR_n1}
\end{figure*}

Figures \ref{fig:comb_AGN_n1} and \ref{fig:comb_PSR_n1} show the calibration plots $\lambda_{rec}$ vs. $\lambda_{inj}$ for all AGN combined and all PSR combined respectively. For PSR, note that the scale is not the same for $n=1$ and $n=2$. This leads to an apparent higher value of the uncertainty. 

\section{Author contributions}\label{sec:annexC}

Initially created by A. Jacholkowska and M. Martinez, the task force was lead by M. Martinez for the MAGIC Collaboration, A. N. Otte for the VERITAS Collaboration and J. Bolmont for the H.E.S.S. Collaboration. J.~Bolmont acted as the main task force leader after the passing of A. Jacholkowska in 2018. He was also the principal coordinator for the writing of the present article.

Software development activities were shared between S. Caroff, A. Gent, D. Kerszberg, C. Levy, T. Lin, L. Nogu\'es, C. Perennes and T. Terzi\'c. M. Ronco contributed studying J\&P and DSR redshift dependence of the time delays, as well as in the writing of the introduction. J. Bolmont, S. Caroff, M. Gaug, A. Gent, D. Kerszberg, C. Levy, T. Lin, T. Terzi\'c provided the IRFs used in the paper. S. Caroff and C. Levy were responsible for producing the final results and plots. Finally, all the authors had a significant contribution writing and editing the draft.


\begin{thebibliography}{}




\bibitem[Aad et al.(2016)]{ATLAS2016} Aad, G., Abbott, B., Abdallah, J., et al. (ATLAS Collaboration) 2016, J. High Energ. Phys. 2016, 26,  \url{https://arxiv.org/abs/1512.02586}
\bibitem[Abdalla et al.(2019)]{Abdalla2019} Abdalla, H., et al. (H.E.S.S. Collaboration) 2019, \apj, 870, 93, 
\url{https://arxiv.org/abs/1901.05209}
\bibitem[Abramowski et al.(2011)]{Abramowski2011} Abramowski, A., et al. (H.E.S.S. Collaboration) 2011, Astropart. Phys., 34, 738, 
\url{https://arxiv.org/abs/1101.3650}
\bibitem[Abramowski et al.(2015)]{Abramowski2015} Abramowski, A., et al. (H.E.S.S. Collaboration) 2015, \apj, 802, 65, \url{https://arxiv.org/abs/1501.05087}
\bibitem[Acciari et al.(2019a)]{Acciari2019a} Acciari, V.~A., et al. (MAGIC Collaboration) 2019a, Nature, 575, 455, \url{https://arxiv.org/abs/2006.07249}
\bibitem[Acciari et al.(2019b)]{Acciari2019b} Acciari, V.~A., et al. (MAGIC Collaboration) 2019b, Nature, 575, 459, \url{https://arxiv.org/abs/2006.07251}
\bibitem[Acciari et al.(2020)]{Acciari2020} Acciari, V.~A., et al. (MAGIC Collaboration) 2020, \prl, 125, 021301, \url{https://arxiv.org/abs/2001.09728}
\bibitem[Acharya et al.(2019)]{Acharya2019} Acharya, B.~S., et al. (CTA Consortium) 2019, World Scientific Publishing Co. Pte. Ltd., \url{https://arxiv.org/abs/1709.07997}
\bibitem[Addazi et al.(2021)]{Addazi2021} Addazi, A., et al. (COST Action CA18108), Prog. Part. Nucl. Phys., in press, \url{https://arxiv.org/abs/2111.05659}
\bibitem[Ade et al.(2016)]{Ade2016} Ade, P. A. R., et al. (Planck Collaboration) 2016, \aap, 594, A13, \url{https://arxiv.org/abs/1502.01589}
\bibitem[Aghanim et al. (2018)]{Planck:2018vyg}
Aghanim, N., et al. (Planck Collaboration) 2020, \aap, 641, A6, \url{https://arxiv.org/abs/1807.06209},
[erratum: \aap, 652 (2021), C4]
\bibitem[Aharonian et al.(2006)]{Aharonian2006} Aharonian, F., et al. (H.E.S.S. Collaboration) 2006, \aap, 457, 899, \url{https://arxiv.org/abs/astro-ph/0607333}
\bibitem[Aharonian et al.(2007)]{Aharonian2007} Aharonian, F., et al. (H.E.S.S. Collaboration) 2007, \apjl, 664, 71, \url{https://arxiv.org/abs/0706.0797}
\bibitem[Ahnen et al.(2017)]{Ahnen2017} Ahnen, M.~L., et al. (MAGIC Collaboration) 2017, \apjs, 232, 9, \url{https://arxiv.org/abs/1709.00346}
\bibitem[Ajello et al.(2019)]{Ajello2019} Ajello, M., et al. (Fermi-LAT Collaboration) 2019, \apj, 878, 52, \url{https://arxiv.org/abs/1906.11403}
\bibitem[Albert et al.(2007)]{Albert2007} Albert, J., et al. (MAGIC Collaboration) 2007, \apj, 669, 862, \url{https://arxiv.org/abs/astro-ph/0702008}
\bibitem[Aleksi\'c et al.(2012)]{Aleksic2012} Aleksi\'c J., et al. (MAGIC Collaboration) 2012, Astropart. Phys., 35, 435, \url{https://arxiv.org/abs/1108.1477}
\bibitem[Aleksi\'c et al.(2016)]{Aleksic2016} Aleksi\'c J., et al. (MAGIC Collaboration) 2016, Astropart. Phys., 72, 76, \url{https://arxiv.org/abs/1409.5594}
\bibitem[Aleksi\'c et al.(2015)]{Aleksic2015} Aleksi\'c J., et al. (MAGIC Collaboration) 2015, JHEAp, 5, 30, \url{https://arxiv.org/abs/1406.6892}
\bibitem[Alfaro et al.(1999)]{Urrutia1999} Alfaro, J. 1999, \prl, 84, 2318, \url{https://arxiv.org/abs/hep-th/0412295}
\bibitem[Ambj{\o}rn et al.(2012)]{Loll2012} Ambj{\o}rn, J., et al. 2012, Phys. Rep., 519, 127, \url{https://arxiv.org/abs/1203.3591}
\bibitem[Amelino-Camelia et al.(1998)]{Amelino1998} Amelino-Camelia, G., et al. 1998, \nat, 393, 763, \url{https://arxiv.org/abs/astro-ph/9712103}
\bibitem[Amelino-Camelia \& Majid(2000)]{Amelino2000} Amelino-Camelia, G., \& Majid, S. 2000, Int. J. Mod. Phys. A, 15, 4301, \url{https://arxiv.org/abs/hep-th/9907110}
\bibitem[Amelino-Camelia(2002)]{Amelino2002} Amelino-Camelia, G. 2002, Nature, 418, 34, \url{https://arxiv.org/abs/gr-qc/0207049}
\bibitem[Amelino-Camelia(2013)]{Amelino2013} Amelino-Camelia, G. 2013, Living Rev. Relativ., 16, 5, \url{https://link.springer.com/article/10.12942/lrr-2013-5}
\bibitem[Amelino-Camelia et al.(2017)]{Ronco2017} Amelino-Camelia, G., Calcagni, G., Ronco, M. 2017, Phys. Lett. B, 774, 630, \url{https://arxiv.org/abs/1705.04876}
\bibitem[Amelino-Camelia et al.(2021)]{Amelino2021} Amelino-Camelia, G., Rosati, G., Bedi\'{c}, S. 2021, Phys. Lett. B, 820, 136595, \url{https://arxiv.org/abs/2012.07790}
\bibitem[Antoniadis et al.(1998)]{Arkani1998} Antoniadis, I., et al. 1998, Phys. Lett. B, 436, 257, \url{https://arxiv.org/abs/hep-ph/9804398}
\bibitem[Barcaroli et al.(2016)]{Barcaroli2016}  Barcaroli, L., et al. 2016, \prd, 95, 024036, \url{https://arxiv.org/abs/1612.01390}
\bibitem[Barrau et al.(2015)]{Barrau2015} Barrau, A., et al. 2015, JCAP,  05, 051, \url{https://arxiv.org/abs/1404.1018}
\bibitem[Barres de Almeida \& Daniel(2012)]{Barres2012} Barres de Almeida, U., \& Daniel, M. K. 2012, Astropart. Phys., 35, 850, \url{https://arxiv.org/abs/1204.2205}
\bibitem[Bernardini et al.(2017)]{Bernardini2017} Bernardini, M.~G., et al. 2017, \aap, 607, A121, \url{https://arxiv.org/abs/1710.08432}
\bibitem[Biteau \& Williams(2015)]{Biteau2015} Biteau, J., \& Williams, D. A. 2015, \apj, 812, 60, \url{https://arxiv.org/abs/1502.04166}
\bibitem[Bolmont et al.(2008)]{Bolmont2008} Bolmont, J., et al. 2008, \apj, 676, 532, \url{https://arxiv.org/abs/astro-ph/0603725}
\bibitem[Bombelli et al.(1987)]{Bombelli1987} Bombelli, L., et al. 1987,  \prl, 59, 521
\bibitem[Brahma \& Ronco(2018)]{Ronco2018} Brahma, S., \& Ronco, M. 2018, Phys. Lett. B,  778, 184, \url{https://arxiv.org/abs/1801.09417}
\bibitem[Brun \& Rademakers(1997)]{ROOT} Brun, R., \& Rademakers, F. 1997, Nucl. Inst. \& Meth. in Phys. Res. A, 389, 81
\bibitem[Calcagni (2017)]{Calcagni2017} Calcagni, G. 2017, J. High Energ. Phys. 2017, 138, \url{https://arxiv.org/abs/1612.05632}
\bibitem[Calcagni et al.(2019)]{Calcagni2019} Calcagni, G., et al. 2019, JCAP, 10, 012, \url{https://arxiv.org/abs/1907.02489}
\bibitem[Caraveo et al.(2001)]{Caraveo2001} Caraveo, P.A., et al. 2001, \apj, 561, 930, \url{https://arxiv.org/abs/astro-ph/0107282}
\bibitem[Carroll et al.(2001)]{Carroll2001}
{Carroll}, S.~M., et al. 2001, \prl, 87, 141601, \url{https://arxiv.org/abs/hep-th/0105082}
\bibitem[Castro-Tirado et al.(2019)]{Castro-Tirado2019} Castro-Tirado, A.~J., et al. 2019, GCN Circulars 23708, \url{https://gcn.gsfc.nasa.gov/gcn3/23708.gcn3}
\bibitem[Chretien et al.(2015)]{Chretien2015} Chretien, M., et al. 2015, Proceedings of ICRC 2015, \url{https://arxiv.org/abs/1509.03545}
\bibitem[Chretien(2015)]{ChretienPhD} Chretien, M. 2015, PhD thesis, \url{https://tel.archives-ouvertes.fr/tel-01271357}
\bibitem[Ellis et al.(2006)]{Ellis2006} Ellis, J., et al. 2006, Astropart. Phys., 25, 402, \url{https://arxiv.org/abs/astro-ph/0510172}
\bibitem[Ellis et al.(2008)]{Ellis2008} Ellis, J., et al. 2008, Astropart. Phys., 29, 158, \url{https://arxiv.org/abs/0712.2781}
\bibitem[Ellis et al.(2019)]{Ellis2019} Ellis, J., et al. 2019, \prd, 99, 083009, \url{https://arxiv.org/abs/1807.00189}
\bibitem[Gambini \& Pullin(1999)]{Gamibini1999} Gambini, R., \& Pullin, J. 1999, \prd, 59, 124021, \url{https://arxiv.org/abs/gr-qc/9809038}
\bibitem[Ganguly et al.(2013)]{Ganguly2013} Ganguly, R., et al. 2013, MNRAS, 435, 1233, \url{https://arxiv.org/abs/1307.6799}
\bibitem[G\"otz et al.(2014)]{Gotz2014} G\"otz, D., et al. 2014, MNRAS, 444, 2776, \url{https://arxiv.org/abs/1408.4121}
\bibitem[Gropp et al.(2019)]{Gropp2019} Gropp, J. D., et al. 2019, GCN Circulars 23688, \url{https://gcn.gsfc.nasa.gov/gcn3/23688.gcn3}
\bibitem[Jacob \& Piran(2008)]{Jacob2008} Jacob, U., \& Piran, T. 2008, JCAP, 01, 031, \url{https://arxiv.org/abs/0712.2170}
\bibitem[Jacobson et al.(2006)]{Liberati2006} Jacobson, T., et al. 2006, Annals Phys.,  321, 150, \url{https://arxiv.org/abs/astro-ph/0505267}
\bibitem[Kaplan et al.(2008)]{Kaplan2008} Kaplan, D.L., et al., 2008, \apj, 677, 1201, \url{https://arxiv.org/abs/0801.1142}
\bibitem[Kosteleck\'y \& Mewes(2008)]{Kostelecky2008} Kosteleck\'y, V.A., \& Mewes, M. 2008, \apj, 689, L1–L4, \url{https://arxiv.org/abs/0809.2846}
\bibitem[Kowalski-Glikman \& Nowak (2002)]{Kowalski2002} Kowalski-Glikma, J., \& Nowak, S. 2002, Phys. Lett. B,  539,  126, \url{https://arxiv.org/abs/hep-th/0203040}
\bibitem[Levy et al.(2021)]{Levy2021} Levy, C., Sol, H., Bolmont, J. 2021, Proceedings of ICRC 2021, \url{https://arxiv.org/abs/2110.06734}
\bibitem[Liberati (2013)]{Liberati2013} Liberati, S. 2013, Class. Quant. Grav., 30, 133001, \url{https://arxiv.org/abs/1304.5795}
\bibitem[Livine et al.(2011)]{Livine2011} Livine, E., et al. 2011, Class. Quant. Grav., 28, 245010, \url{https://arxiv.org/abs/1104.5509}
\bibitem[Magueijo  \& Smolin(2004)]{Smolin2004} Magueijo , J., \& Smolin, L. 2004, Class. Quant. Grav.,  21, 1725, \url{https://arxiv.org/abs/gr-qc/0305055}
\bibitem[Mao(2011)]{Mao2011} Mao, L.S., 2011, New Astronomy, 16, 503
\bibitem[Martinez \& Errando(2009)]{Martinez2009} Mart{\'{\i}}nez, M. \& Errando, M. 2009, Astropart. Phys., 31, 226, \url{https://arxiv.org/abs/0803.2120}
\bibitem[Mattingly (2005)]{Mattingly2005} Mattingly, D. 2005, Living Rev. Rel., 8, 5, \url{https://link.springer.com/article/10.12942/lrr-2005-5}
\bibitem[Mavromatos(2010)]{Mavromatos2010} Mavromatos, N. E. 2010, IJMPA, 25, 5409, \url{https://arxiv.org/abs/1010.5354}
\bibitem[Niedermaier et al.(2006)]{Reuter2006}  Niedermaier, M., et al. 2006, Living Rev. Relat., 9, 5, \url{https://link.springer.com/article/10.12942/lrr-2006-5}
\bibitem[Oriti(2001)]{Oriti2006} Oriti, D. 2001, Rept. Prog. Phys., 64 1703, \url{https://arxiv.org/abs/gr-qc/0106091}
\bibitem[Perennes et al.(2020)]{Perennes2020} Perennes, C. , Sol, H. and Bolmont, J. 2020, \aap, 633, A143, \url{https://arxiv.org/abs/1911.10377}
\bibitem[Pfeifer (2018)]{Pfeifer2018}  Pfeifer, C.  2018, Phys. Lett. B, 780, 246, \url{https://arxiv.org/abs/1802.00058}
\bibitem[Polyakov (1981)]{Polyakov1981} Polyakov, A. M. 1981,  Phys. Lett. B, 103, 207
\bibitem[Pueschel(2019)]{Pueschel2019} Pueschel, E. 2019, Proceedings of ICRC 2019, \url{https://arxiv.org/abs/1908.04163}
\bibitem[Rosati et al.(2015)]{Rosati2015} Rosati, G., et al. 2015, \prd, 92,  124042, \url{https://arxiv.org/abs/1507.02056}
\bibitem[Rovelli(2007)]{Rovelli2007} Rovelli, C. 2007,  Cambridge University Press, Cambridge U.K.
\bibitem[Selsing et al.(2019)]{Selsing2019} Selsing, J., et al. 2019, GCN Circulars 23695, \url{https://gcn.gsfc.nasa.gov/gcn3/23695.gcn3}
\bibitem[Terzi\'c et al.(2021)]{terzic2021} Terzi\'c. T., Kerszberg, D., Stri\v{s}kovi\'c, J. 2021, Universe, 7, 345, \url{https://arxiv.org/abs/2109.09072}
\bibitem[Vasileiou et al.(2013)]{Vasileiou2013} Vasileiou, V., et al. 2013, \prd, 87, 122001, \url{https://arxiv.org/abs/1305.3463}
\bibitem[Vasileiou et al.(2015)]{Vasileiou2015} Vasileiou, V., et al. 2015, \nat, 11, 344
\bibitem[Xu \& Ma (2016a)]{XuMa2016a} Xu, H., \& Ma, B.-Q. 2016, Astropart. Phys., 82, 72, \url{https://arxiv.org/abs/1607.03203}
\bibitem[Xu \& Ma (2016b)]{XuMa2016b} Xu, H., \& Ma, B.-Q. 2016, Phys. Lett. B, 760, 602, \url{https://arxiv.org/abs/1607.08043}
\bibitem[Xu \& Ma (2018)]{XuMa2018} Xu, H., \& Ma, B.-Q. 2018, JCAP, 1801, 050, \url{https://arxiv.org/abs/1801.08084}
\bibitem[Zitzer et al.(2013)]{Zitzer2013} Zitzer, B., et al. 2013, Proceedings of ICRC 2013, \url{https://arxiv.org/abs/1307.8382}
\bibitem[Zyla et al.(2020)]{Zyla:2020zbs}
Zyla, P.A., et al. (Particle Data Group) 2020, Prog. Theor. Exp. Phys. 2020, 083C01, \url{https://pdg.lbl.gov/rpp-archive/index-2021.html}


\end{thebibliography}
\end{document}